\documentclass[lettersize,journal]{IEEEtran}
\usepackage{amsmath,amsfonts,bm}
\usepackage{algorithm}
\usepackage{array}
\usepackage[caption=false,font=normalsize,labelfont=sf,textfont=sf]{subfig}
\usepackage{textcomp}
\usepackage{stfloats}
\usepackage{url}
\usepackage{verbatim}
\usepackage{graphicx}
\usepackage{cite}
\usepackage{empheq}
\usepackage{algpseudocode}
\newtheorem{remark}{Remark}
\usepackage{booktabs}
\usepackage{siunitx}
\usepackage{makecell}
\usepackage{flushend}
\usepackage{tabularx}
\usepackage{booktabs}
\usepackage{multirow}
\usepackage{xcolor}
\usepackage{tabularx}

\let\originalleft\left
\let\originalright\right
\newcommand{\newleft}{\mathopen{}\mathclose\bgroup\originalleft}
\newcommand{\newright}{\aftergroup\egroup\originalright}

\newcommand{\ra}[1]{\renewcommand{\arraystretch}{#1}}
\DeclareMathOperator*{\argmax}{arg\,max}

\def\ptheta(#1){p_{\param}\newleft(#1\newright)}
\def\pthetac(#1|#2){p_{\param}\newleft(#1\,\vline\,#2\newright)}

\newcommand{\state}{\bm{x}}
\newcommand{\control}{\bm{u}}
\newcommand{\meas}{\bm{z}}
\newcommand{\processnoise}{\bm{w}}
\newcommand{\measnoise}{\bm{v}}
\newcommand{\dyn}{\bm{f}}
\newcommand{\measfun}{\bm{h}}
\newcommand{\param}{\bm{\theta}}
\newcommand{\Qc}{\bm{Q}}
\newcommand{\Rk}{\bm{R}}
\newcommand{\Pk}{\bm{P}}

\begin{document}

\title{Accelerating Bayesian Optimization \\ for Nonlinear State-Space System Identification \\ with Application to Lithium-Ion Batteries}

\author{Hao Tu,~\IEEEmembership{Student Member,~IEEE}, Jackson Fogelquist, Iman Askari, Xinfan Lin,~\IEEEmembership{Senior Member,~IEEE},\\ Yebin Wang,~\IEEEmembership{Senior Member,~IEEE}, Shiguang Deng, Huazhen Fang,~\IEEEmembership{Senior Member,~IEEE}
\thanks{
	
		
		Hao Tu, Iman Askari and Shiguang Deng are with the Department of Mechanical Engineering, University of Kansas, Lawrence, KS 66045 USA (e-mail: tuhao@ku.edu, askari@ku.edu, sdeng@ku.edu).

        Jackson Fogelquist and Xinfan Lin are with the Department of Mechanical and Aerospace Engineering, University of California, Davis, CA 95616 USA
        (e-mail: jbfogelquist@ucdavis.edu, lxflin@ucdavis.edu).

        Yebin Wang is with the Mitsubishi Electric Research Laboratories, Cambridge, MA 02139 USA (e-mail: yebinwang@ieee.org).
        
        Huazhen Fang is with the Department of Mechanical Engineering, Michigan State University, East Lansing, MI 48824 USA (e-mail: hfang@msu.edu).

	}

}



\maketitle

\begin{abstract}
This paper studies system identification for nonlinear state-space models, a problem that arises across many fields yet remains challenging in practice. Focusing on maximum likelihood estimation, we employ Bayesian optimization (BayesOpt) to address this problem by leveraging its derivative-free global search capability  enabled by surrogate modeling of the likelihood function. Despite these advantages, standard BayesOpt often suffers from slow convergence, high computational cost, and practical difficulty in attaining global optima under limited computational budgets, especially for high-dimensional nonlinear models with many unknown parameters. 

To overcome these limitations, we propose an accelerated BayesOpt framework that integrates BayesOpt with the Nelder--Mead method. Heuristics-based, the Nelder--Mead method provides fast local search, thereby assisting BayesOpt when the surrogate model lacks fidelity or when over-exploration occurs in broad parameter spaces. The proposed framework incorporates a principled strategy to coordinate the two methods, effectively combining their complementary strengths. The resulting hybrid approach significantly improves both convergence speed and computational efficiency while maintaining strong global search performance. In addition, we leverage an implicit particle filtering method to enable accurate and efficient likelihood evaluation. We validate the proposed framework on the identification of the BattX model for lithium-ion batteries, which features ten state dimensions, 18 unknown parameters, and strong nonlinearity. Both simulation and experimental results demonstrate the effectiveness of the proposed approach as well as its advantages over alternative methods.
\end{abstract}


\section{Introduction}

State-space models (SSMs), introduced by Rudolf Kalman in~\cite{Kalman:IFAC:1960}, have transformed the way of describing dynamic systems since their inception. Their impact arises from several key strengths, including a clear mathematical formulation, strong representational capacity, and broad applicability to many classes of systems. Today, SSMs have achieved widespread use across scientific, engineering, and economic fields, providing a foundation for system analysis, monitoring, diagnosis, and control~\cite{elliott:book:1995, west:book:1999, cappé:book:2006, doucet:Book:2001}.

To deploy SSMs in real-world applications, it is essential to estimate their parameters from measured data accurately and efficiently. This problem has attracted sustained attention since the 1960s~\cite{Cox:TAC:1964, Ljung:TAC:1979}. To date, maximum likelihood (ML) estimation has become a leading method due to its statistical optimality, natural incorporation of probabilistic noise models, and ease of integration with a range of statistical tools~\cite{KantasL:SS:2015}. Despite these advantages, several nontrivial challenges remain. A core difficulty is that, for an SSM, parameter identification typically requires joint---either explicit or implicit---inference of the latent state, which evolves over time but is not directly observed. This challenge is further exacerbated in the case of nonlinear SSMs, where the nonlinearity complicates the search over both unknown parameters and hidden state trajectories. These challenges have motivated a growing body of research, which we survey below.

\subsection{Literature review} \label{Sec: literature review}
ML estimation provides a principled framework to study the parameter estimation problem for SSMs. The literature has developed the following categories of methods based on this framework. 

{\em 1) Direct gradient-based search.} A natural way to perform ML estimation is to maximize the likelihood function using gradient‑based optimization. For nonlinear SSMs, however, the likelihood takes an intractable form, making its gradients hard to obtain. To address this issue, the work in~\cite{Doucet:Statistics:2003} derives an explicit gradient expression and then approximates it using particle methods, while~\cite{Poyiadjis:ACC:2006} employs stochastic perturbation to estimate the gradients. Another line of work, known as iterative filtering, approximates the gradients through posterior moments conditioned on data~\cite{Ionides:PNAS:2006}. All these approaches rely on particle filtering to enable gradient approximation. A major concern, however, is that the approximate gradients can exhibit high variance. Mitigating this variance typically requires a large number of particles, which significantly increases computational cost.

{\em 2) Expectation-maximization (EM)-based identification.} The EM algorithm performs maximum-likelihood estimation in the presence of latent variables, making it useful for identifying state-latent SSMs. It operates through two iterative steps: an E‑step, which evaluates the expected log-likelihood, and an M‑step, which maximizes this quantity with respect to the model parameters. Several studies have explored the use of EM for SSM identification~\cite{Schon:Auto:2011,Wills:Auto:2013,Picchini:CS:2018}. In general, evaluating the likelihood in the E-step requires state estimation, for which particle filtering and smoothing are often the methods of choice, as shown in~\cite{Schon:Auto:2011,Wills:Auto:2013,Picchini:CS:2018}. However, EM-based approaches often suffer from slow convergence, heavy computation, and strong sensitivity to initialization; these issues become more pronounced when the SSM is strongly nonlinear or high-dimensional. A generalization of EM is variational inference, which provides an alternative way for identifying SSMs. This approach posits a tractable family of distributions for the unknown variables and then selects the member of this family with minimum Kullback-Leibler divergence with respect to the true posterior~\cite{COURTS:Auto:2023}. 

{\em 3) Particle-based Markov chain Monte Carlo (MCMC) methods.} This class of methods combines particle filtering with MCMC to enable SSM identification. The central idea is to embed a particle filter within an MCMC scheme so that each iteration uses particle‑filter‑based likelihood estimates to propose or evaluate parameter updates, while still targeting the correct posterior distribution~\cite{Johansen:SC:2008,Andrieu:JRSS:2010}. Although these methods are, in principle, capable of exploring high-dimensional parameter spaces, they often exhibit slow convergence and substantial computational cost. A few studies have thus appeared to alleviate this issue. Among them, the work in~\cite{PITT:JE:2012} analyzes the optimal number of particles needed for stable convergence and studies the use of a fully adapted auxiliary  particle filter  to reduce computation. The use of multiple Markov chains, rather than a single one, and parallel computing architectures have been explored in~\cite{Mingas:IJAR:2017} to accelerate computation.

While the above-reviewed studies offer useful approaches to nonlinear SSM identification, they often converge slowly, require heavy computation, and easily get stuck in local optima. These limitations frequently lead to failures in parameter estimation when an SSM involves a large number of parameters, yet with little prior knowledge about them. 

To deal with these issues, Bayesian optimization (BayesOpt) has recently emerged as a compelling approach. BayesOpt is a data-driven optimization framework that constructs a surrogate model---typically a Gaussian process (GP)---to approximate the objective function, iteratively refines this surrogate using observed data, and exploits it to guide the search for optimal solutions \cite{frazier:tutorial:2018,Bobak:IEEE:2016}. The GP surrogate is expressive, provides uncertainty quantification, and offers analytical tractability in both objective function evaluation and optimization. These properties enable BayesOpt to handle expensive-to-evaluate objective functions, eliminate the need for gradients in search, balance exploitation of promising regions with exploration of uncertain ones, and search effectively for global optima in complex landscapes. As such, BayesOpt offers a compelling way for nonlinear SSM identification, with several budding studies having  explored this direction. For example, the work in \cite{Dahlin:arXiv:2017} combines BayesOpt with particle filtering to construct a Laplace approximation of the otherwise intractable parameter posterior. However, despite its strengths, BayesOpt can still be computationally demanding, particularly due to its exploration-exploitation tradeoff and the increasingly flat optimization landscapes encountered in high-dimensional spaces. To improve efficiency, the study in \cite{Mahdi:TNNLS:2022} projects the original high-dimensional parameter space onto a reduced subspace before running BayesOpt. Nevertheless, despite these advances, BayesOpt has not yet fully realized its potential for nonlinear SSM identification. Addressing its limitations in scalability and efficiency remains an open challenge, serving as the primary motivation for the work presented in this paper.

\subsection{Review of Battery Model Identification}  \label{Sec: Battery ID}
While   nonlinear SSM identification is needed across many fields, we focus on a case study involving lithium-ion batteries (LiBs) in this paper. LiBs have been a driving force behind the global transition toward electrification and decarbonization, with ever-growing applications in electric mobility, smart grids, and renewable energy systems. Central to the practical use of LiBs are battery management systems (BMSs), which ensure safety, performance, and longevity from cell to system level. BMSs often rely on equivalent circuit models (ECMs)---circuit analogs composed of resistors, capacitors, and voltage sources---to emulate LiBs' behaviors\cite{Plett:2015}. Accurately identifying ECM parameters is therefore a critical task. 

The literature has presented several approaches. A traditional yet popular method is parameter calibration through experiments, where specially designed current profiles are applied to a LiB cell to stimulate  parameter-dependent responses in the measurements~\cite{Lin:JPS:2014,Biju:AE:2023,Ahmed:SAE:2015}. Although straightforward and easy to implement, this approach often provides limited accuracy, restricts itself to fixed current profiles, and often requires long testing durations. Furthermore, this approach is constrained to simple ECMs, as selectively exciting specific parameter-dependent dynamics becomes difficult in more advanced models~\cite{Goshtasbi:JPS:2024}. A more formalized alternative is to estimate ECM parameters by minimizing the prediction error between ECM-based prediction and measurement data, an approach known as the prediction-error framework~\cite{ljung:book:1999}. This generally results in nonlinear optimization problems, e.g.,~\cite{Sitterly:TSE:2011,Ahmed:SAE:2015,Tian:JES:2020}. Another methodology relies on statistical inference, such as ML or maximum a posteriori estimation, which also leads to optimization problems for uncovering unknown parameters~\cite{Tian:TCST:2020}.
To date, few studies have addressed LiB model identification through the lens of SSM identification. To our knowledge, the only study is~\cite{Tian:TCST:2020}, which considers a linear second-order ECM with only a small number of parameters. Recently, several studies have begun using BayesOpt for LiB model identification, though limited to relatively simple models~\cite{Tu:ACC:2024, Pi:MECC24:2024}. In contrast, this work tackles a more ambitious problem: identifying a high-dimensional nonlinear ECM with many parameters,  by enhancing BayesOpt.

\subsection{Statement of Contributions}

To achieve SSM identification, we propose  to integrate BayesOpt with the Nelder--Mead method in this paper. The Nelder--Mead method searches for optima by operating on a simplex and applying sequential geometric transformations that reshape and translate the simplex toward regions of improved objective value~\cite{Nelder:TCJ:1965,Lagarias:SIAM:1998}. Like BayesOpt, it is gradient-free; unlike BayesOpt, it performs local optimization and can rapidly generate new query points through its simplex transformation rules. These properties make the Nelder--Mead method strongly complementary to BayesOpt: it can warm-start BayesOpt and efficiently refine the search once BayesOpt identifies a promising region. This synergy can accelerate the overall search toward global optima in high-dimensional parameter spaces for SSMs.

We conduct a systematic investigation of this idea, leading to the following contributions:

\begin{itemize}

\item We develop an enhanced BayesOpt framework by integrating  BayesOpt with the Nelder--Mead method. This framework furnishes principled switching conditions to coordinate the two methods coherently, so that both play to their respective strengths to accelerate convergence and computation in the parameter search.

\item  As part of the framework, we propose to compute the ML-based objective function for nonlinear SSM identification using the unscented implicit particle filter (U-IPF) introduced in~\cite{ASKARI:AUTO:2022}. The U-IPF method achieves accurate state estimation using a small number of particles concentrated in high-probability regions of the posterior, thereby offering substantial computational savings and further accelerating the overall parameter search.

\item We apply the proposed framework to identify the BattX model, an ECM for LiBs recently developed in~\cite{Biju:AE:2023}. The model is nonlinear and contains 18 unknown parameters and ten state variables.  Both simulation and experimental studies  validate the performance of the proposed approach in tackling this challenging identification problem.

\end{itemize}

A preliminary version of this work appeared in~\cite{Tu:ACC:2024}, focusing on applying BayesOpt to identify a temperature-dependent ECM. Compared to~\cite{Tu:ACC:2024}, the present paper provides substantial extensions, including:  1) the development of an accelerated BayesOpt methodology, 2) the formulation of likelihood evaluation via U-IPF, 3) validation using the BattX model, which is of considerably higher order and involves many more parameters than the model considered in~\cite{Tu:ACC:2024}, and 4) validation based on experimental data.

\subsection{Organization of the paper}
The remainder of the paper is organized as follows. Section~\ref{sec: problem formulation} presents the nonlinear SSM identification problem from the perspective of ML estimation. Section~\ref{sec: accBayesOpt} proposes the accelerated BayesOpt framework. Section~\ref{sec: Likelihood Evaluation} presents the method of likelihood evaluation via the U-IPF method. Section V validates the proposed approach by applying it to identify the BattX model for LiBs using both simulation and experimental data. Finally, Section~\ref{sec: Conclusion} concludes the paper.

\section{Problem Formulation} \label{sec: problem formulation}
We consider a general-form nonlinear SSM:
\begin{subequations}\label{Eqn: SS model}
\begin{empheq}[left=\empheqlbrace]{align}
\label{Eqn: dynamic model}
\state_{k+1} &= \dyn\!\left(\state_{k}, \control_{k}, \param\right) + \processnoise_{k},\\
\meas_k  &= \measfun\!\left(\state_k, \control_{k}, \param\right) + \measnoise_{k},
\end{empheq}
\end{subequations}
where $\state_k \in \mathbb{R}^{n_{\state}}$ is the state, $\control_k \in \mathbb{R}^{n_{\control}}$ is the input, $\meas_k \in \mathbb{R}^{n_{\meas}}$ is the measurement. Further, $\dyn$ and $\measfun$ are the state transition and measurement functions, respectively, which are both parametrized in the unknown parameter vector $\param \in \mathbb{R}^{n_{\param}}$. Besides, $\processnoise_k$ and $\measnoise_k$ are white Gaussian noise processes, with $\processnoise_k \sim \mathcal{N}(\bm{0},\Qc)$ and $\measnoise_k \sim \mathcal{N}(\bm{0},\Rk)$. For brevity, we omit $\control_k$ in subsequent expressions as it is known.

Given the SSM in~\eqref{Eqn: SS model}, we aim to extract $\param$ from the measurement dataset $\meas_{1:T} = \{ \meas_1, ..., \meas_T\}$ using ML estimation:
\begin{align} \label{Eqn: ML}
    \param^* = \argmax_{\param} L(\param) \coloneqq \log \ptheta(\meas_{1:T}),
\end{align}
where $\ptheta(\meas_{1:T})$ is the likelihood distribution of $\meas_{1:T}$. Using Bayes' rule, the log-likelihood function $L(\param)$ can be rewritten as
\begin{align} \label{eqn: Ltheta}
L(\param) = \log \ptheta(\meas_{1}) + \sum_{k=2}^{T} \log \pthetac(\meas_{k} | \meas_{1:k-1}).
\end{align}
Here, $\pthetac(\meas_{k} | \meas_{1:k-1})$ satisfies
\begin{align} \label{Eqn: yutheta}
    \pthetac(\meas_{k} | \meas_{1:k-1}) &= \int \pthetac(\meas_{k}|\state_{k})\pthetac(\state_{k}| \meas_{1:k-1}) \, d\state_k.
\end{align}
Using the Markovian property of the SSM and Bayes' rule, we have the following recursion governing the evolution of $p_{\param}(\state_k|\meas_{1:k-1})$:

\begin{subequations}\label{Eqn: Recursion}
\begin{align}
p_{\param}(\state_k|\meas_{1:k-1})
&=
\int p_{\param}(\state_k|\state_{k-1})
\, p_{\param}(\state_{k-1}|\meas_{1:k-1})
\, d\state_{k-1},
\label{Eqn: prediction}
\\
p_{\param}(\state_k|\meas_{1:k})
&=
\frac{
p_{\param}(\meas_k|\state_k)\,
p_{\param}(\state_k|\meas_{1:k-1})
}{
p_{\param}(\meas_k|\meas_{1:k-1})
}.
\label{Eqn: update}
\end{align}
\end{subequations}

The problem in~\eqref{Eqn: ML} is nontrivial. A primary challenge lies in the intractability of $L(\param)$, as no closed-form solution to \eqref{Eqn: yutheta}--\eqref{Eqn: Recursion} is available for the nonlinear SSM. As such, existing approaches must rely on gradient methods, EM algorithms, or MCMC sampling, as reviewed in Section~\ref{Sec: literature review}, each of which entails substantial approximations. These approximations not only increase computational burden but also limit effectiveness in parameter search. The difficulty escalates further when the parameter dimension $n_{\param}$ is large. To address these challenges, we next develop a novel BayesOpt-based framework for solving the system identification problem.








\begin{remark}
The above ML-based problem formulation readily extends to the case of multiple datasets. Suppose there are $M$ separate datasets, $\meas^{(m)}_{1:T_m} = \{\meas^{(m)}_1,\ldots,\meas^{(m)}_{T_m}\}$,
$m=1,\ldots,M$, are available.  Then, the total log-likelihood is given by the sum of the individual log-likelihoods, i.e.,
\begin{align*}
        \log p_{\param}\!\left(\meas^{(1)}_{1:T_1},\ldots,\meas^{(M)}_{1:T_M}\right) = \sum_{m=1}^{M} \log p_{\param}\!\left(\meas^{(m)}_{1:T_m}\right).
    \end{align*}
This property allows the proposed approach to be directly applied to multiple datasets collected on the same system to improve identification accuracy, as often needed in practical applications.

\end{remark}

\section{Accelerated BayesOpt for SSM Identification}\label{sec: accBayesOpt}

This section presents an improved BayesOpt framework for nonlinear SSM identification. We begin by reviewing the standard BayesOpt approach, then introduce the Nelder–Mead method, and finally describe how to combine the two to enable more effective and efficient parameter search.

\subsection{BayesOpt}
BayesOpt considers $L(\param)$ to be a black-box function and uses a GP to capture probabilistic relations between $\param$ and $L(\param)$. The GP thus serves as a surrogate model to approximate $L(\param)$; based on this surrogate, an acquisition function selects the next evaluation point for $\param$. This procedure iteratively updates the GP with new samples to improve the surrogate accuracy, continuing the search until convergence toward a global optimum. For clarity, we denote the surrogate as $\hat{L}(\param)$ to distinguish it from the original objective $L(\param)$.
As a GP, $\hat L(\param)$ takes the following prior distribution:
\begin{align*}
    \hat{L}(\param) \sim \ \mathcal{GP}(\mu(\param), k(\param, \param')),
\end{align*}
where $\mu(\cdot)$ and $k(\cdot,\cdot)$ are the mean and kernel functions, respectively. Note that $k(\param, \param')$ encodes the correlation between $\hat{L}(\param)$ and $\hat{L}(\param')$, where both $\param$ and $\param'$ belong to the same parameter space.  Common choices of $k(\cdot,\cdot)$ include the squared exponential kernel and Mat{\'e}rn kernel~\cite{frazier:tutorial:2018}. Also, both $\mu(\cdot)$ and $k(\cdot,\cdot)$ contain hyperparameters that are adapted using data obtained from evaluating $L(\param)$ during the search. This adaptation process is commonly referred to as GP training. Suppose that $L(\param)$ has been evaluated at $q$ distinct parameter points, and denote the resulting evaluation data as $\left\{ ( \param_i, \mathcal{L}_i), i=1,\ldots q \right\}$, where $\mathcal{L}_i = L(\param_i)$. Then, the posterior distribution of $\hat{L}(\param)$ can be obtained as
\begin{align} \label{Eqn: GP posterior}
    \hat{L}(\param) \; | \; \mathcal{L}_{1:q} \sim \mathcal{N} (\mu_q(\param), \Sigma_q(\param)),
\end{align}
where 
\begin{align*}
    \mu_q(\param) &= \mu(\param) + \bar{k}(\param,\param_{1:q})K^{-1}\left(\mathcal{L}_{1:q}-\mu(\param_{1:q})\right), \\
    \Sigma_q(\param) &= k(\param,\param)-\bar{k}(\param,\param_{1:q})K^{-1}\bar{k}(\param)^\top.
\end{align*}
Here,
\begin{align*}
    \mathcal{L}_{1:q} &=
    \begin{bmatrix}
         \mathcal{L}_{1}\ \cdots\ \mathcal{L}_{q}
    \end{bmatrix}^\top, \\
    \mu(\param_{1:q}) &= 
    \begin{bmatrix}
        \mu(\param_1)\ \cdots\ \mu(\param_q)
    \end{bmatrix}^\top, \\
    \bar{k}(\param,\param_{1:q}) &= 
    \begin{bmatrix}
        k(\param, \param_1) & \cdots & k(\param, \param_q)
    \end{bmatrix}, \\
    K &=  
    \begin{bmatrix}
         k(\param_1, \param_1) & \cdots & k(\param_1, \param_q) \\
         \vdots & \ddots & \vdots \\
         k(\param_1, \param_q)^\top & \cdots & k(\param_q, \param_q)
    \end{bmatrix}.
\end{align*}
The posterior distribution in~(\ref{Eqn: GP posterior}) represents the prediction of $\hat{L}(\param)$ at an arbitrary $\param$ based on the existing evaluation data. It is needed in determining the next query point $\param_{q+1}$.

BayesOpt uses the so-called acquisition function to guide the search for $\param_{q+1}$. For the acquisition function design, a popular choice is the expected improvement. The improvement means the increase of $\hat{L}(\param)$ with respect to the maximum of the so-far observed $L(\param_{1:q})$. As $\hat{L}(\param)$ is probabilistic, we must consider the expectation of the improvement. Specifically, denoting $L^* = \max \mathcal L_{1:q}$, the expected improvement is defined as
\begin{align*}
    \mathrm{EI}(\param \mid \mathcal L_{1:q}) = \mathbb{E} \Big[ \Big( \hat{L}({\param}) - L^* \Big)^+ \; \Big| \; \mathcal L_{1:q}  \Big],
\end{align*}
where $(\cdot)^+ = \max(\cdot,0)$, and the expectation is taken over the posterior distribution given by~(\ref{Eqn: GP posterior}). Naturally, $\param_{q+1}$ is selected to be the point that maximizes  $\mathrm{EI}(\param \mid \mathcal L_{1:q})$:
\begin{align*}
    \param_{q+1} = \argmax_{\param} \mathrm{EI}(\param \mid \mathcal L_{1:q}).
\end{align*}
Such a way to search for $\param_{1:q}$ will not only exploit the available knowledge embodied by $\mathcal L_{1:q}$, but also allow to explore the parameter space by harnessing the probabilistic uncertainty. This balance between exploitation and exploration facilitates the search for global optima.

\subsection{The Nelder--Mead Method}

\begin{figure}[t!]
    \centering
    \includegraphics[width = .4\textwidth,trim={6.5cm 2.3cm 6.4cm 5.8cm},clip]{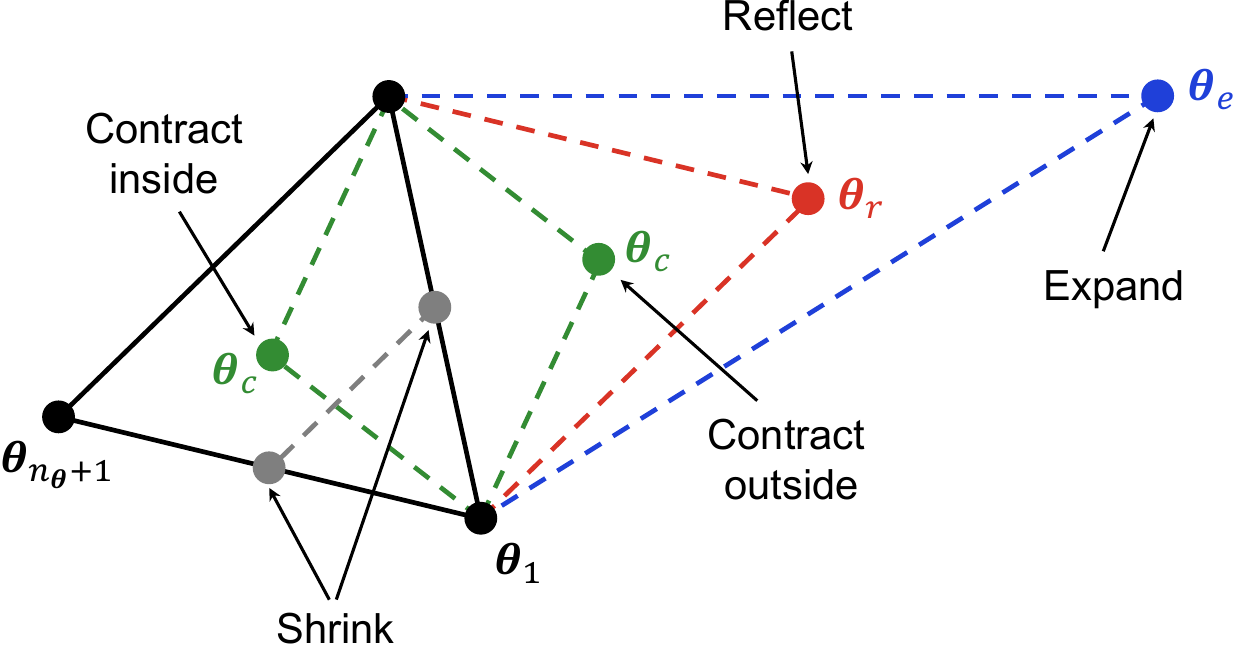}
    \caption{Illustration of the Nelder--Mead method, adapted from~\cite{Cheng:CBC:2015}.}
    \label{Fig: N-M optimization}
\end{figure}

The Nelder--Mead method is a derivative-free heuristic algorithm for minimizing an objective function. The method maintains a simplex, which is a polytope formed by $n+1$ points in an $n$-dimensional space (e.g., a triangle in 2D and a tetrahedron in 3D), and iteratively adapts the simplex geometry using only function evaluations to move toward regions of lower objective values~\cite{Nelder:TCJ:1965,Lagarias:SIAM:1998}. 

Consider the task of maximizing $L(\param)$, or equivalently minimizing $L^-(\param):=-L(\param)$. At each iteration, the method deals with a simplex based on $\{(\param_i, L(\param_i)), i=1, \ldots, n_{\param}+1\}$, which are the evaluations of $L(\param)$ at different points. The Nelder--Mead method proceeds as below.

\begin{enumerate}

\item Order the points such that
\[
L^-(\param_1) \le L^-(\param_2) \le \cdots \le L^-(\param_{n_{\param}+1}).
\]

\item Compute the centroid of all points except the worst one by 
\[
\param_o=\frac{1}{n_{\param}}\sum_{i=1}^{n_{\param}} \param_i,
\]
which serves as a reference for subsequent geometric transformations.

\item {\em Reflection.} Reflect the worst point $\param_{n_{\param}+1}$ through the centroid $\param_o$ via
\[
\param_r = \param_o + \alpha(\param_o - \param_{n_{\param}+1}),
\quad \alpha>0.
\]
\begin{enumerate}

\item If
$
L^-(\param_1) \le L^-(\param_r) < L^-(\param_n),
$
replace the worst point $\param_{n_{\param}+1}$ by $\param_r$, and go to Step 1).

\item {\em Expansion.} If
$
L^-(\param_r)<L^-(\param_1),
$
which implies the reflection yields an improvement, expand the search to find the next query point by
\[
\param_e=\param_o+\gamma(\param_r-\param_o), \] 
where $\gamma>1$; replace the worst point $\param_{n_{\param}+1}$ by $\param_e$ if $\param_e$ is better than $\param_r$, i.e.,
$
L^-(\param_e) < L^-(\param_r),
$
and by $\param_r$ if otherwise; then go to Step 1).

\item {\em Contraction.} If
$
L^-(\param_r)>L^-(\param_n),
$
the reflection brings no improvement, thus contract the current simplex via either outside contraction
\[
\param_c = \param_o+\rho(\param_r-\param_o),
\]
if
$
L^-(\param_r)<L^-(\param_{n_{\param}+1}),
$
or inside contraction
\[
\param_c = \param_o+\rho(\param_{n_{\param}+1}-\param_o),
\]
if
$
L^-(\param_r) \ge L^-(\param_{n_{\param}+1}),
$
where $0<\rho<1$; replace the worst point $\param_{n_{\param}+1}$ by $\param_c$ if $\param_c$ is better than $\param_r$, i.e.,
$
L^-(\param_c) < L^-(\param_r),
$
and go to Step 4) if otherwise.

\end{enumerate}

\item {\em Shrinkage} If all the above steps fail, shrink the simplex toward the best point $\param_1$ by
\[
\param_i = \param_1+\sigma(\param_i-\param_1), \quad i=2, \ldots, n_{\param}+1,
\]
with $0<\sigma<1$, and go to Step 1). 

\end{enumerate}

Fig.~\ref{Fig: N-M optimization} illustrates the Nelder--Mead method. Through successive reflection, expansion, contraction, and shrinkage operations, the Nelder--Mead method adaptively reshapes the simplex to explore regions of decreasing objective values. The method requires no gradient information, making it suitable for optimization problems with non-analytical cost functions and with unavailable or expensive-to-evaluate gradients. Although the method does not guarantee convergence to global minima, it often shows good empirical performance on various problems.

\subsection{Accelerating BayesOpt with the Nelder--Mead Method}

\begin{figure*}[t!]
    \centering
    \includegraphics[width = \textwidth,trim={.4cm 6.75cm .3cm 8.8cm},clip]{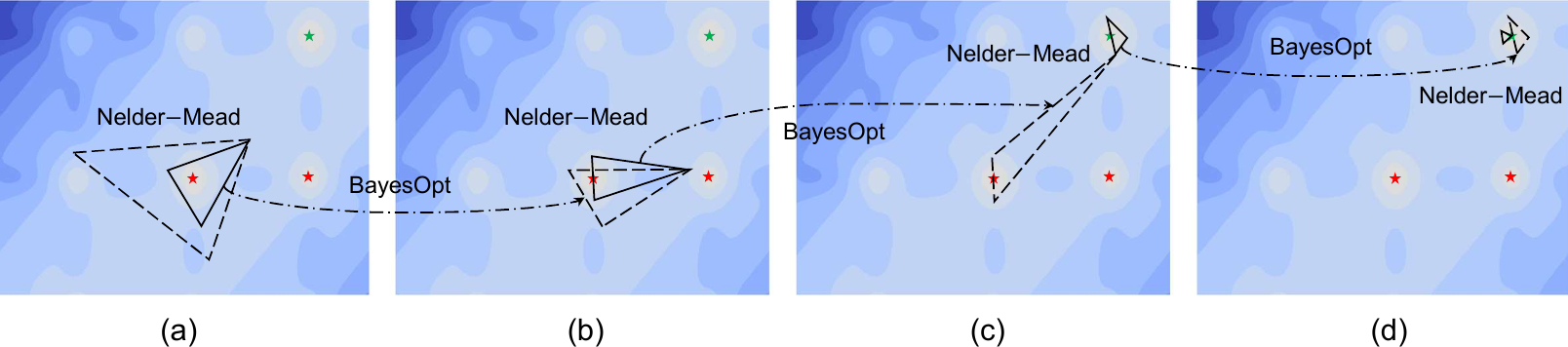}
    \caption{Illustration of the accelerated BayesOpt approach. The contours show the landscape of $L(\param)$. The green and red stars mark the global and local   optima, respectively. The dashed and solid triangular simplexes correspond to the initial and final simplexes of each Nelder--Mead run. During optimization, BayesOpt guides the search toward the global optimum, while Nelder--Mead performs rapid local refinement. Across successive runs, the simplexes gradually shrink, indicating that Nelder--Mead progressively narrows its search region.}
    \label{Fig:hybrid search}
\end{figure*}

\begin{table}
\centering
\caption{Comparison of BayesOpt, Nelder--Mead optimization and the accelerated BayesOpt algorithms.}
\renewcommand{\arraystretch}{1.5}

\begin{tabular}{l | c c c}
\toprule
 & BayesOpt & Nelder--Mead & \makecell{Accelerated\\ BayesOpt}\\  
\midrule
Search strategy & Surrogate-driven & Heuristic & Mix of both \\  
\hline
Global optimality & Yes & No & Yes \\  
\hline
\makecell[l]{Computational \\ complexity} & High & Low & \makecell{Lower than \\ standard \\ BayesOpt} \\  
\hline
\makecell[l]{Robustness to \\ initialization} & High & Low & High \\  
\bottomrule
\end{tabular}

\label{Table: BO&NM comparison}
\end{table}

What draws our attention is the similarities and differences between BayesOpt and the Nelder–Mead method. Both algorithms are designed to address optimization problems in which the objective function is black-box or expensive to evaluate, or lacks analytically tractable gradients. Implementation-wise, both methods seek to identify and drive the search toward promising regions of the parameter space using function evaluations. Despite these commonalities, the two approaches differ in their underlying mechanisms. BayesOpt constructs a GP–based surrogate model to encode the knowledge learned from past evaluations and exploits this surrogate to guide future queries. Due to the surrogate's probabilistic nature, BayesOpt's search balances exploitation of promising regions and exploration of uncertain regions, thus providing a theoretical capability for global optimization. However, BayesOpt often incurs substantial computational overhead and slow convergence, because of repeated surrogate updates and acquisition optimization. Practically limited computing budgets can also compromise its actual convergence to global optima.  In contrast, the Nelder–Mead method relies solely on function evaluations and conducts the search by adaptively reshaping the simplex geometry. This results in significantly lower computational cost and much faster search, albeit with guarantees limited to local optimality at most. Table~\ref{Table: BO&NM comparison} sums up the comparison between the two approaches. 

\begin{algorithm}[t!]
\caption{Accelerated BayesOpt for nonlinear SSM identification}
\label{alg:accBO}
\begin{algorithmic}[1]

\vspace{3pt}

\Statex \quad {\color{gray}\textit{--- Initialization ---}}

\vspace{3pt}

\State Draw $D$ parameter samples $\{\param_i\}_{i=1}^{D}$
\State Evaluate $L(\param_i)$ using Algorithm~\ref{alg:likelihood}
\State Set up the observation pool 
$
\mathcal P=\{(\param_i,L(\param_i))\}_{i=1}^{D}
$


\Repeat

\vspace{3pt}

\Statex \quad {\color{gray}\textit{--- Nelder--Mead local search ---}}

\vspace{3pt}

\Repeat
\State Perform simplex transformations to search
\State Evaluate the likelihood at the new query points
\State Add the new observations to $\mathcal P$
\Until{The average distance among the vertices of the simplex $d<d_{\mathrm{lim}}$ for the present round \textbf{or} no improvement for $p$ iterations}

\State Update the GP surrogate using $\mathcal P$

\vspace{3pt}

\Statex \quad {\color{gray}\textit{--- BayesOpt global search ---}}

\vspace{3pt}

\Repeat

\State Perform expectation-improvement-based search
\State Evaluate the likelihood at new query points
\State Add the new observation to $\mathcal P$
\State Update the GP surrogate using $\mathcal P$
\Until{The most recent observation ranks among the top $m$ values in $\mathcal P$
\textbf{or} BayesOpt stagnates for $s$ iterations}


\Until{BayesOpt stagnates for $s$ iterations}

\vspace{3pt}

\Statex \quad {\color{gray}\textit{--- Final local refinement ---}}

\vspace{3pt}

\State Initialize Nelder--Mead with the best $n_{\param}+1$ points in $\mathcal P$

\Repeat
\State Run Nelder--Mead iterations and evaluate $L(\param)$
\Until{$d<d_{\mathrm{final}}$}

\State \textbf{return}
$
\param^\ast=\argmax_{\param_i\in\mathcal P}L(\param)
$

\end{algorithmic}
\end{algorithm}

These observations highlight that the Nelder–Mead method is complementary to BayesOpt to accelerate the latter toward more efficient search, less computation, and potentially better global convergence. To bring this idea into fruition, we propose the following framework to fuse the two methods across different stages of a parameter search. 

\begin{itemize}

\item {\em Beginning stage.} BayesOpt often shows low efficiency in the early stage of the search. This is because a limited number of evaluation points are available initially to train the GP surrogate, and in turn, the low-informative GP may misguide the selection of subsequent query points. Thus, the Nelder–Mead method provides an effective alternative in this stage, by rapidly and heuristically exploring the parameter space to supply informative samples for BayesOpt.

\item {\em Medium stage.} The search in this stage alternates between BayesOpt and the Nelder--Mead method so that they mutually benefit each other. Specifically, when BayesOpt identifies a promising region, the Nelder--Mead method then kicks in to perform fast local refinement within that region; when the local search is exhaustive, BayesOpt resumes, updating the surrogate with the latest evaluations obtained by the Nelder--Mead method and then running subsequent exploitive/exploratory steps. 

\item {\em Final stage.} As the search draws to a close, BayesOpt often continues to probe the vicinity around the global optimum or even unpromising regions, rather than settling, partly due to its tendency for over-exploration~\cite{NEURIPS:David:2019}. The Nelder--Mead method then can take over to efficiently run the final iterations and promote convergence through focused search near the global optimum. 

\end{itemize}

To implement the above framework,  it is critical to set up the switching and termination conditions between BayesOpt and the Nelder--Mead method. Our proposal is as follows.

\begin{itemize}

\item {\em Switch from BayesOpt to Nelder--Mead.} Enable the switching when BayesOpt identifies a candidate point whose objective value is better than that of the top $m \ (m < n_{\param} + 1)$ points within the current pool of evaluated samples. Upon switching, add this newly identified point to the pool; then, select the best $m$ points from the updated pool, and randomly pick $n_{\param}-m+1$ points from the rest of the pool to construct the simplex for the next Nelder--Mead run.

\item {\em Switch from Nelder--Mead to BayesOpt.} Enable the switching when the average distance $d$ among the vertices of the simplex is smaller than the preset limit $d_{\mathrm{lim}}$ within this round of run or when the Nelder--Mead method yields no improvement within $p$ consecutive iterations. Here,   $d_{\mathrm{lim}}$ should decrease across successive rounds to narrow down the Nelder--Mead method's search range. In this study,   $d_{\mathrm{lim}}$ is reduced by half at each round relative to the distance of the initial simplex. Upon switching, add all newly evaluated points to the pool, on which BayesOpt updates its surrogate for the next search. 

\item {\em Switch from BayesOpt to Nelder--Mead in the final stage.} Enable the switching when BayesOpt yields no improvement after exceeding a preset maximum $s$ number of iterations. Then, execute the Nelder--Mead method initialized with a simplex formed by the best $n_{\param}+1$ points in the pool.

\item {\em Termination.} Terminate the entire search when the average distance $d$ of the vertices of the Nelder--Mead method's simplex falls below a small preset tolerance $d_{\mathrm{final}}$ in the final stage, indicating convergence. 

\end{itemize}

Fig.~\ref{Fig:hybrid search} provides an illustration for the proposed accelerated BayesOpt approach. Algorithm~\ref{alg:accBO} summarizes the procedures for the approach. This approach preserves the key strengths of standard BayesOpt, including its ability to perform global search, while providing new advantages, like fast search, reduced computation, and improved robustness to initialization, due to incorporating the Nelder--Mead method.  This  hybrid approach can substantially enhance the efficiency in search and computation, as well as estimation accuracy, for SSM identification with high-dimensional parameter sets.

\section{Likelihood Evaluation via U-IPF}\label{sec: Likelihood Evaluation}

As shown in Section~\ref{sec: accBayesOpt}, the accelerated BayesOpt algorithm requires evaluating the likelihood function $L(\param)$ during the optimization process. This task boils down to computing $\pthetac(\meas_{k} | \meas_{1:{k-1}})$, as indicated by \eqref{eqn: Ltheta}. While $\pthetac(\meas_{k} | \meas_{1:{k-1}})$ has no closed form, we can leverage particle filtering to approximate it. This approach generates a set of particles with importance weights $\{{\state}_{k|k-1}^i, w_{k-1}^i\}_{i=1}^{N_p}$ to form an empirical distribution of 
\begin{align*}
\pthetac(\state_{k} | \meas_{1:{k-1}}) \approx \sum_{i=1}^{N_p} w_{k-1}^i\delta(\state_{k}-{\state}_{k|k-1}^i).  
\end{align*}
Following~\eqref{Eqn: yutheta}, $\pthetac(\meas_{k} | \meas_{1:{k-1}})$ can then be approximated as
\begin{align}\label{eqn: yuthetaappox}
    \pthetac(\meas_{k} | \meas_{1:{k-1}}) \approx \sum_{i=1}^{N_p} w_{k-1}^i\pthetac(\meas_{k}|{\state}_{k|k-1}^i),
\end{align}
which is used to evaluate $L(\param)$ in \eqref{eqn: Ltheta}. 
Some studies, e.g., \cite{Schon:Auto:2011}, have used the bootstrap particle filter to generate the particles. However, this method typically requires a large number of particles to adequately cover the support of the posterior state distribution in order to achieve sufficient estimation accuracy, and the required particle number also grows rapidly with the SSM's dimension. This causes   hefty computation and   risks of filter collapse. One can use the auxiliary particle filter (APF) to mitigate this issue~\cite{Pitt:JASA:1999,Pitt:Warwick:2002}, but the extent to which it helps is still limited for nonlinear SSMs.

To address this limitation, we propose to use the U-IPF method developed in~\cite{ASKARI:AUTO:2022} here. This method traces to the implicit particle filtering framework in~\cite{Chorin:Math:2010}, which shows that significantly fewer particles are needed when they concentrate in high-probability regions of the posterior state distribution. The U-IPF method exploits the unscented transform and Kalman update to compute the particles; structurally, it resembles a bank of unscented Kalman filters   operating in parallel. As shown in~\cite{ASKARI:AUTO:2022}, this method offers both high sampling efficiency and estimation accuracy even for high-dimensional systems. 

In its implementation, the U-IPF method   computes three ensembles recursively: $\{{\state}_{k|k-1}^i, w_{k-1}^i\}_{i=1}^{N_p}$,   $\{{\meas}_{k|k-1}^i, w_{k-1}^i\}_{i=1}^{N_p}$, and $\{{\state}_{k|k }^i, w_{k}^i\}_{i=1}^{N_p}$, which approximate $\pthetac(\state_{k} | \meas_{1:{k-1}})$, $\pthetac(\meas_{k} | \meas_{1:{k-1}})$, and $\pthetac(\state_{k} | \meas_{1:{k}})$, respectively. 
The method runs as below.   
\begin{itemize}

\item \textit{Particle prediction (time update).} 
At time $k-1$, there exists an ensemble   $\{\state_{k-1 | k-1}^i, w_{k-1}^i\}_{i=1}^{N_p}$. Within the U-IPF framework, each particle $\state_{k-1}^i$ has an associated covariance $\Pk_{k-1|k-1}^i$, and  is propagated forward   through the unscented transform ($\mathsf{UT}$)~\cite{Sarkka:Cambridge:2023}:
\begin{subequations}
\begin{align}\label{Eqn: UT-dyn-1}
\left[{\state}_{k|k-1}^i,\; \hat{\Pk}_{\state, k|k-1}^i \right]
&=
\mathsf{UT}\!\left(
\dyn,\;
\state_{k-1 | k-1}^i,\; 
\Pk_{k-1 | k-1}^i
\right), \\ \label{Eqn: UT-dyn-2}
{\Pk}_{k|k-1}^i
&=
\hat{\Pk}_{\state,k|k-1}^i + \Qc.
\end{align}
\end{subequations} 
Similarly,    we perform 
\begin{subequations}
\begin{align}\label{Eqn: UT-meas-1} \nonumber 
\left[ {\meas}_{k|k-1}^i,\;
\hat{\Pk}_{\meas,k|k-1}^i,\;
{\Pk}_{\state\meas,k|k-1}^i \right]
&=  \\
&\hspace{-4em} \mathsf{UT}\!\left(
\measfun,\;
{\state}_{k|k-1}^i,{\Pk}_{k|k-1}^i
\right), \\
\label{Eqn: UT-meas-2}
{\Pk}_{\meas,k|k-1}^i
&= \hat{\Pk}_{\meas,k|k-1}^i + \Rk.
\end{align}
\end{subequations}

\item \textit{Particle update (measurement update).} 
Given the measurement $\meas_k$,  ${\state}_{k|k-1}^i$ undergoes the Kalman-type update:
\begin{subequations}
\begin{align}
\label{eqn:mean}
    \tilde{\state}_{k|k}^i &= {\state}_{k|k-1}^i
    + {\Pk}_{\state\meas,k|k-1}^i({\Pk}_{\meas,k|k-1}^i)^{-1}(\meas_k-{\meas}_{k | k-1}^i),\\
\label{eqn:cov}
    \Pk_{k|k}^i &= {\Pk}_{k|k-1}^i
    - {\Pk}_{\state\meas,k|k-1}^i({\Pk}_{\meas,k|k-1}^i)^{-1}({\Pk}_{\state\meas,k|k-1}^i)^\top .
\end{align}
\end{subequations} 

\item \textit{Implicit  sampling and weighting.} 
A new particle is   sampled via
\begin{align} \label{eqn:particleUIPF}
\state_{k|k}^i = \tilde{\state}_{k|k-1}^i + \sqrt{\Pk_k^i}\,\boldsymbol{\xi}^i,
\end{align}
where $\boldsymbol{\xi}^i \sim \mathcal{N}(\bm{0},\alpha \mathbf{I})$ with $0< \alpha \ll1$.
The particle weight $w_k^i$ is computed via
\begin{align}\label{eqn:weightUIPF}
w_k^i = {w_{k-1}^i\, \pthetac(\meas_k|\state_{k-1}^i) \over \sum_{j=1}^{N_p} w_{k-1}^j\, \pthetac(\meas_k|\state_{k-1}^j)}. 
\end{align}



\end{itemize}
The above outlines the procedure for implementing the U‑IPF method; interested readers are referred to \cite{ASKARI:AUTO:2022} for full details. By design, this method focuses on identifying highly probable samples and requires only a small number of them to achieve efficient and accurate computation. After executing the U‑IPF,  $L(\param)$ can be evaluated directly according to \eqref{eqn: Ltheta}.

\begin{algorithm}[t!]
\caption{Evaluation of $L(\param)$ via U-IPF}
\label{alg:likelihood}
\begin{algorithmic}[1]

\State Initialize an ensemble of particles with weights and covariances 
$\{ \state_{0|0}^i, w_{0|0}^i, \Pk_{0|0}^i \}_{i=1}^{N_p}$
according to $p(\state_{0})$ 

\State Set $k \gets 1$ 

\Repeat

\State Compute   
$\{ {\state}_{k|k-1}^i  \}_{i=1}^{N_p}$ via~\eqref{Eqn: UT-dyn-1}--\eqref{Eqn: UT-dyn-2}

\State Compute approximately  $\pthetac(\meas_{k}|\meas_{1:k-1})$
via \eqref{eqn: yuthetaappox} 

\State Compute 
$\{{\meas}_{k|k-1}^i \}_{i=1}^{N_p}$
via \eqref{Eqn: UT-meas-1}--\eqref{Eqn: UT-meas-2} 

\State Compute
$\{\tilde{\state}_{k|k }^i \}_{i=1}^{N_p}$ by 
(\ref{eqn:mean})--(\ref{eqn:cov}) 

\State Generate 
$\{\state_{k|k}^i, w_{k}^i\}_{i=1}^{N_p}$
via
(\ref{eqn:particleUIPF})--(\ref{eqn:weightUIPF}) 

\State Do resampling if necessary 

\State $k \gets k+1$.

\Until{$k > T$}

\State Evaluate $L(\param)$ via \eqref{eqn: Ltheta} 

\end{algorithmic}
\end{algorithm}

\begin{remark} 
For an SSM with deterministic dynamics, there is no need to run particle filtering; instead, one can directly generate state trajectory samples and then evaluate   $L(\param)$ via~\eqref{eqn: yuthetaappox}. In this case, computation can be  substantially reduced, and additional details  are available in~\cite{Tu:ACC:2024}.
\end{remark}

\section{Case Study: Application to System Identification for LiBs}\label{sec: LiB identification}

\begin{figure*}[t!]
    \centering
    \includegraphics[width = 0.95\textwidth,trim={7.8cm 6.8cm 7cm 11cm},clip]{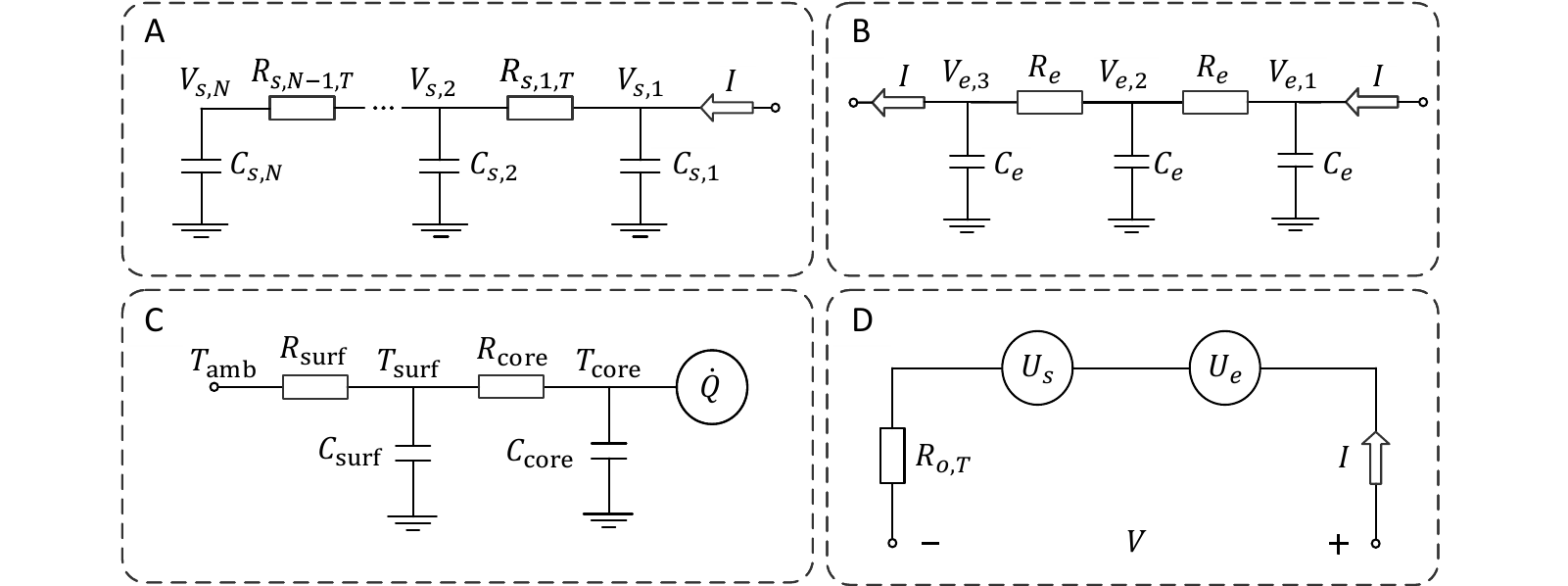}
    \caption{The BattX model comprising four coupled sub-circuits.}
    \label{Fig:BattX}
\end{figure*}

In this section, we present a case study that uses the accelerated BayesOpt approach to enable system identification for LiBs. As reviewed in Section~\ref{Sec: Battery ID}, the growing demand for accurate models in LiB-powered applications has spurred increasing research interest in LiB system identification. While several methods are available for relatively simple LiB models, a clear gap remains for more sophisticated models that aim to capture LiB behaviors in greater fidelity. An example in case is the BattX model proposed in~\cite{Biju:AE:2023}. We therefore apply the proposed BayesOpt approach to this model using both simulation and experimental data. The obtained results validate the effectiveness of our approach.

\subsection{Identification Problem for the BattX model}
 
The BattX model is the first ECM developed to predict the behavior of a LiB cell operating across a wide range of C-rates, from low to high, motivated by emerging applications like electric aircraft. As shown in Fig.~\ref{Fig:BattX}, the model comprises multiple coupled sub-circuits, each designed to approximate a major physical process within the cell. Specifically, sub-circuits A and B emulate lithium-ion diffusion in the solid electrode and the electrolyte, respectively; sub-circuit C captures the thermal dynamics during charge and discharge; and sub-circuit D predicts the terminal voltage response.  By design, this ECM is physically interpretable, and its simple structure ensures computational efficiency. A detailed presentation of the model's governing equations is provided in the Appendix. 

When formulated as an SSM, we discretize the continuous-time dynamics using the Runge--Kutta method and incorporate additive process and measurement noise terms. This procedure yields a discrete-time model of the form given in \eqref{Eqn: SS model}. The model's state, input, and measurement vectors are summarized as
\begin{align*}
\state &= [V_{s,1}\ \cdots \ V_{s,N}\ V_{e,1} \ V_{e,2} \ V_{e,3}\ T_\mathrm{core}\ T_\mathrm{surf}]^\top \in \mathbb{R}^{N+5} , \\
\control &= [I\ T_\mathrm{amb}]^\top \in \mathbb{R}^{2}, \\
\meas &= [V\ T_\mathrm{surf}]^\top \in \mathbb{R}^{2},
\end{align*}
where each variable is illustrated in Fig.~\ref{Fig:BattX}. The model parameters include electrical and thermal resistances and capacitances, as well as coefficients appearing in parameterized functions that capture the dependence of certain variables or parameters on others. A summary of these parameters, totaling 18 in number, is provided in Table~\ref{Table: Sim result}. For the sake of parameter identifiability and with physical soundness, we assume the following: 1) $C_{s,i} = \eta_i C_{s,1},\ R_{s,i} = \sigma_i R_{s,1}$, where $\eta_i$ and $\sigma_i$ are pre-specified ratios, 2) the SoC-OCV function is available, as is usual in the literature, and 3) the statistics of the process and measurement noises are known.

\subsection{Simulation-Based Identification}\label{sec: numerical simulation}

\begin{table}[t!]\centering
\ra{1.2}
\caption{Parameters involved in the BattX model, their nominal values, search ranges in identification, and identified values.}
 \begin{tabular}{ l | l l l }
\toprule
Parameters & Search range  & Identified & True \\
\midrule
$C_{s,1}\ [\si{F}]$ & 2000$\sim$5000  & 4435.352 & 4521 \\
$R_{s,1}\ [\si{\Omega}]$ & 0$\sim$0.5  & 0.1148 & 0.114 \\
$C_{e}\ [\si{F}]$ & 0$\sim$5000  & 3327.462 & 3691 \\
$R_{e}\ [\si{\Omega}]$ & 0$\sim$0.1  & 0.0086 & 0.007 \\
$C_{\mathrm{core}}\ [\si{J/K}]$ & 0$\sim$100  & 36.807 & 40 \\
$C_{\mathrm{surf}}\ [\si{J/K}]$ & 0$\sim$50  & 12.181 & 10 \\
$R_{\mathrm{core}}\ [\si{K/W}]$ & 0$\sim$10  & 1.935 & 2 \\
$R_{\mathrm{surf}}\ [\si{K/W}]$ & 0$\sim$10  & 3.233 & 3 \\
$\beta_1$ & 0$\sim$1  & 0.6960 & 0.789 \\
$\beta_2$ & 0$\sim$1  & 0.6460 & 0.317 \\
$\gamma_1\ $ & 0$\sim$0.1  & 0.0463 & 0.046 \\
$\gamma_2\ $ & -0.1$\sim$0  & -0.0399 & -0.035 \\
$\gamma_3\ $ & 0$\sim$0.1  & 0.0287 & 0.029 \\
$\kappa_1$ & 0$\sim$100  & 34.292 & 30 \\
$\kappa_2$ & 0$\sim$100  & 82.437 & 70 \\
$c_1\ $ & -0.001$\sim$0  & -0.00022 & -0.0004 \\
$c_2\ $ & 0$\sim$0.01  & 0.0019 & 0.002 \\
$c_3\ $ & -0.01$\sim$0  & -0.0011 & -0.001 \\
\bottomrule
\end{tabular}
\label{Table: Sim result}
\end{table}

\begin{figure}[t!]
    \centering
    \subfloat[]{
        \includegraphics[width=0.48\textwidth,trim={4cm 8.4cm 4cm 8.5cm},clip]{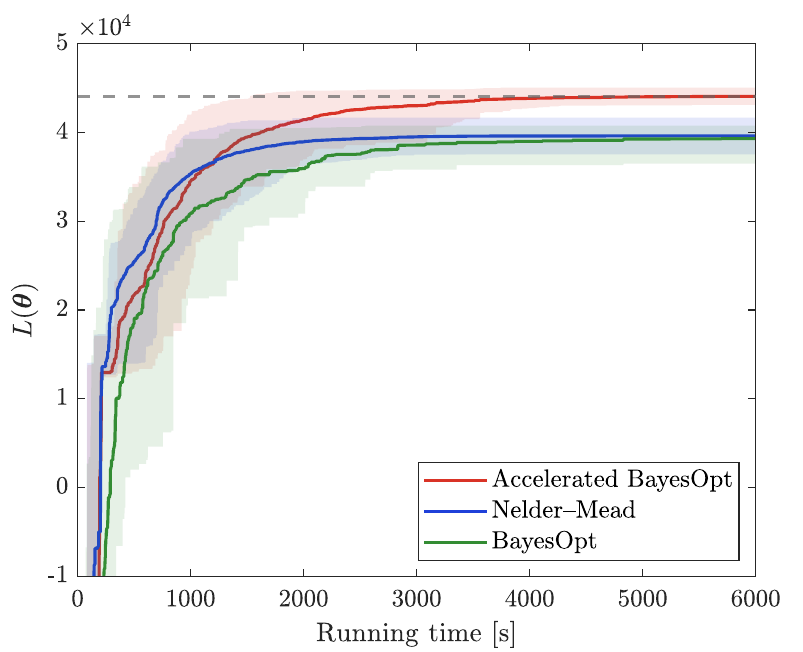}
        \label{Fig:AlgComp_a}
    }
    \hfill
    \subfloat[]{
        \includegraphics[width=0.48\textwidth,trim={4cm 8.4cm 4cm 8.5cm},clip]{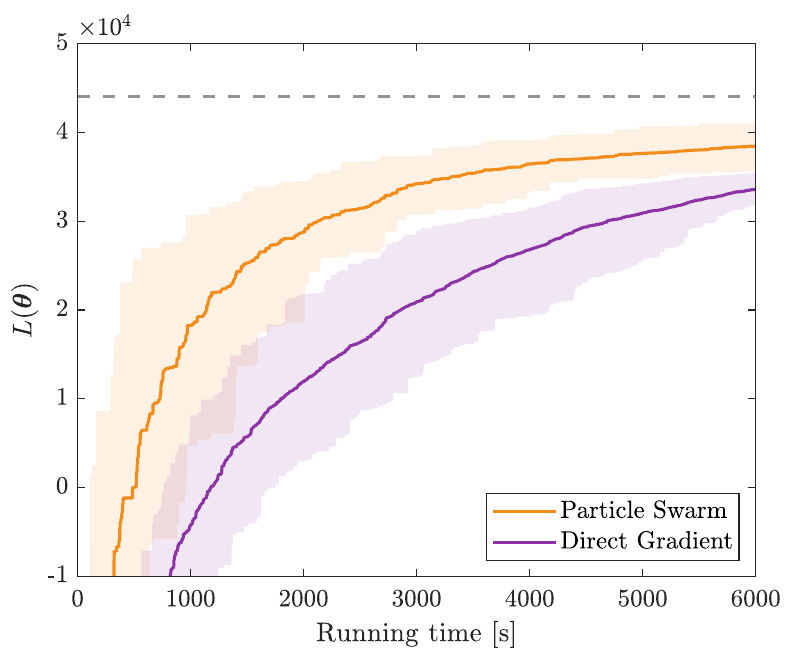}
        \label{Fig:AlgComp_b}
    }
    
    \caption{Convergence performance comparison over 50 independent runs. 
The mean log-likelihood and its variability across runs are shown. 
The dashed black line indicates the log-likelihood evaluated at the nominal parameter $\param^*$. 
All methods use the U-IPF-based likelihood evaluator with $N_p=100$ particles. 
}
    
    \label{Fig:AlgComp}
\end{figure}

\begin{figure}[t!]
    \centering
    
    \subfloat[]{
    \centering
    \includegraphics[width = 0.24\textwidth,trim={.7cm .2cm 8.5cm .4cm},clip]{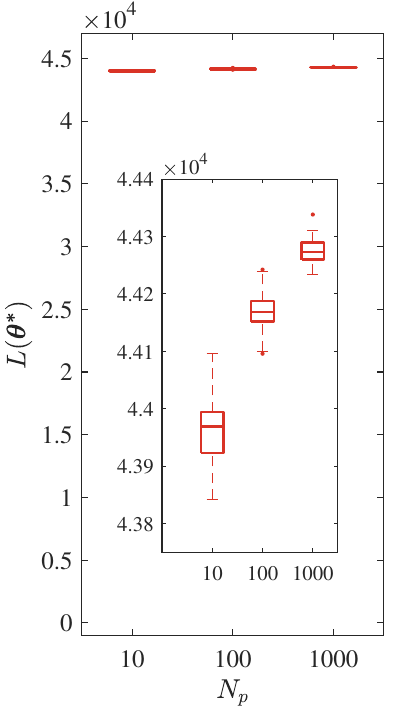}}
    \subfloat[]{
    \centering
    \includegraphics[width = 0.24\textwidth,trim={8.6cm .2cm .6cm .4cm},clip]{PFcomp_new-eps-converted-to.pdf}}
    
    \caption{Comparison of log-likelihood evaluations using (a) U‑IPF and (b) APF around the nominal $\param^*$ with $N_p$ particles. Each filter is run independently 1,000 times for each value of $N_p$.}
    \label{Fig:Likelihood comp}
\end{figure}

\begin{figure*}[t!]
\centering
    \subfloat{
    \centering
    \includegraphics[width = .24\textwidth,trim={.2cm 0cm .9cm .3cm},clip]{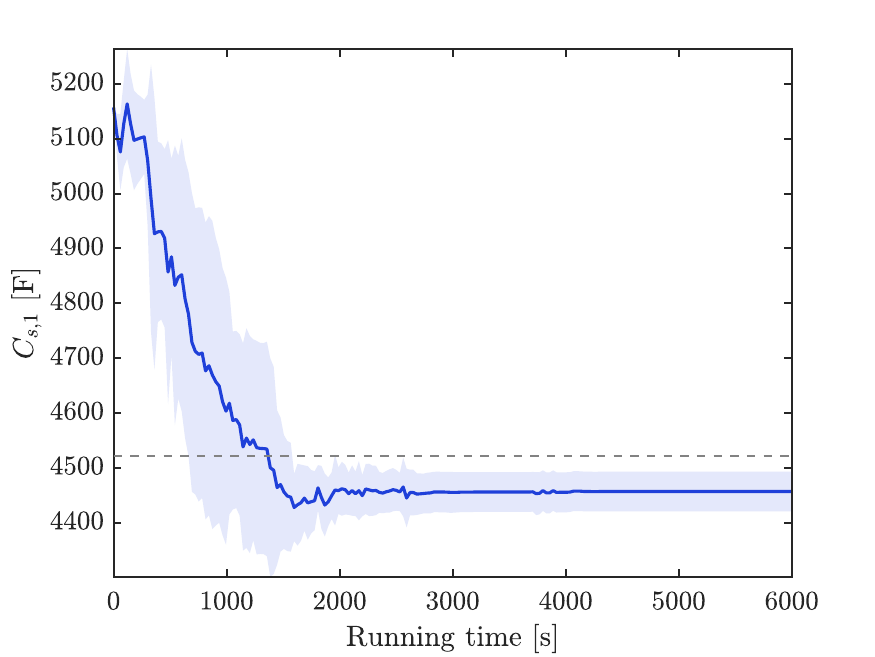}}
    \subfloat{
    \centering
    \includegraphics[width = .24\textwidth,trim={.2cm 0cm .9cm .3cm},clip]{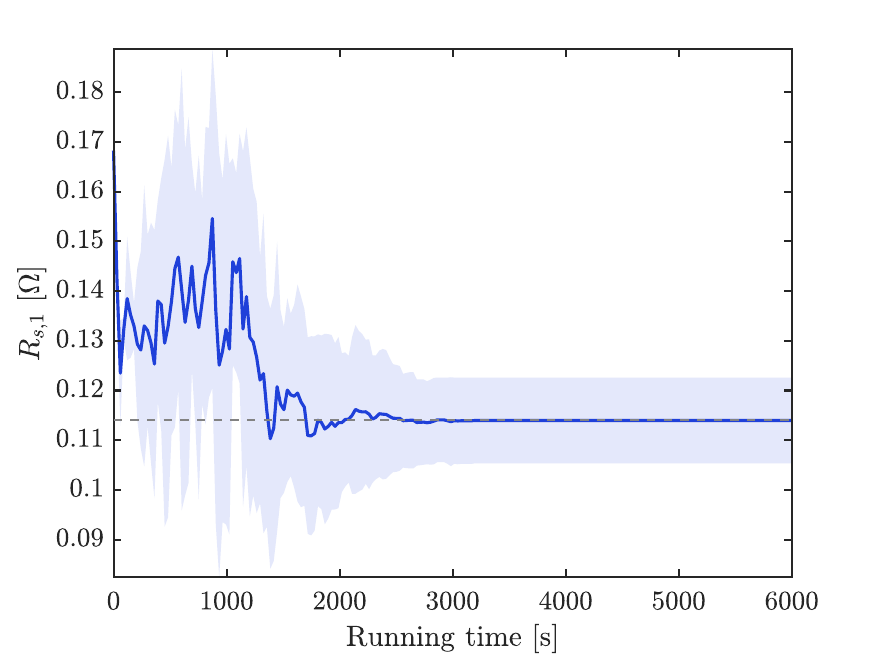}}
   \subfloat{
    \centering
    \includegraphics[width = .24\textwidth,trim={.2cm 0cm .9cm .3cm},clip]{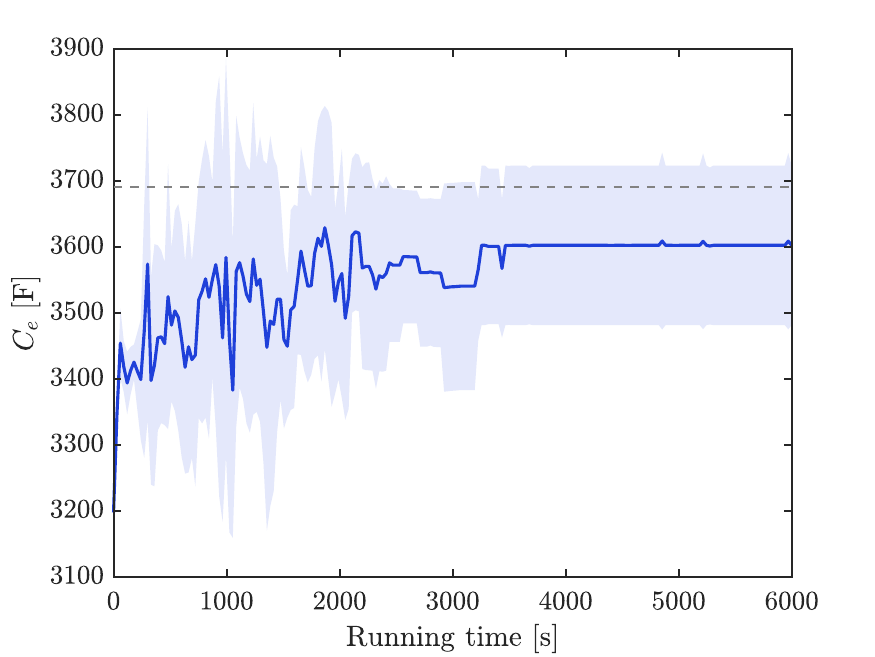}}
    \subfloat{
    \centering
    \includegraphics[width = .24\textwidth,trim={.2cm 0cm .9cm .3cm},clip]{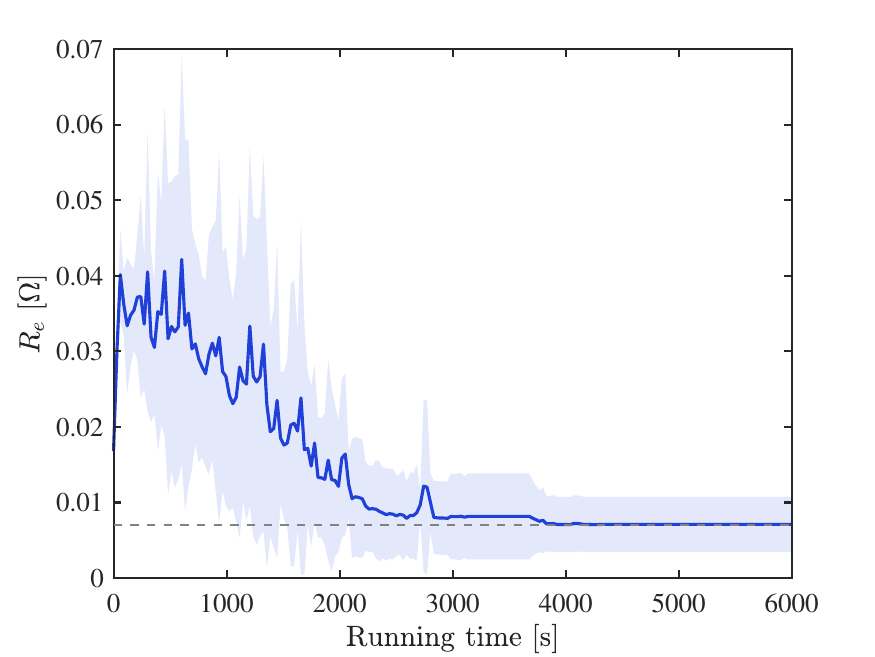}}

    \subfloat{
    \centering
    \includegraphics[width = .24\textwidth,trim={.2cm 0cm .9cm .3cm},clip]{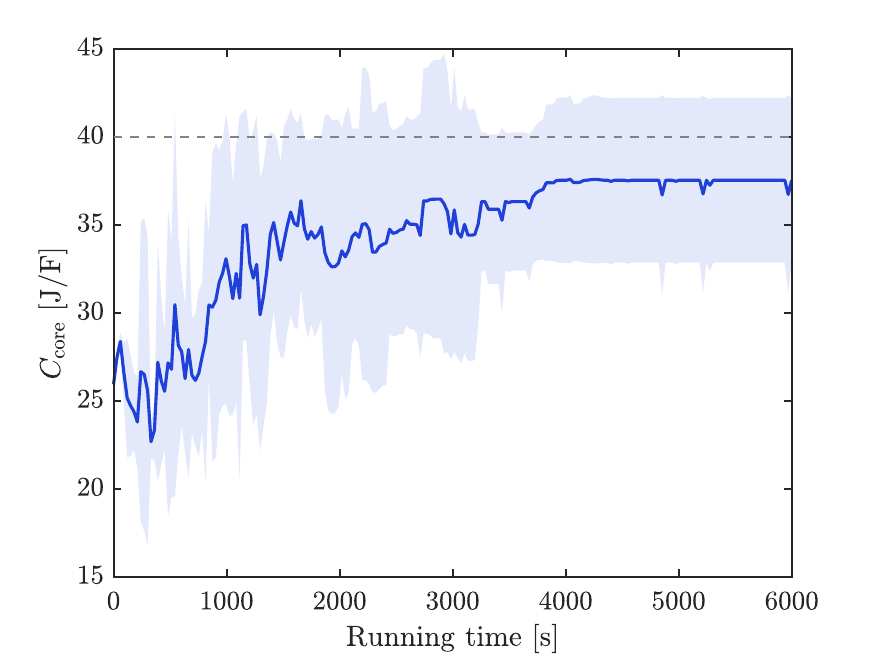}}
    \subfloat{
    \centering
    \includegraphics[width = .24\textwidth,trim={.2cm 0cm .9cm .3cm},clip]{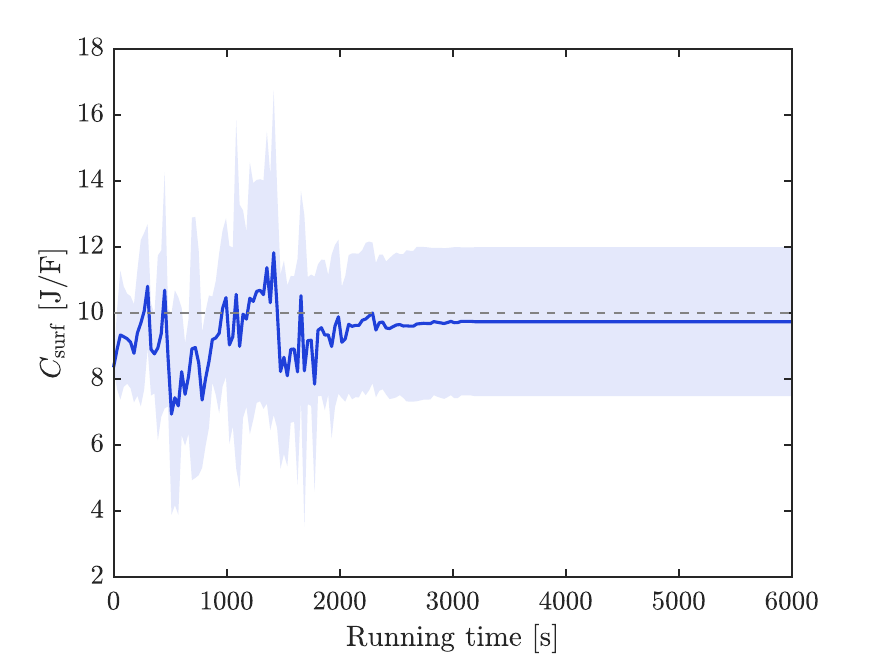}}
    \subfloat{
    \centering
    \includegraphics[width = .24\textwidth,trim={.2cm 0cm .9cm .3cm},clip]{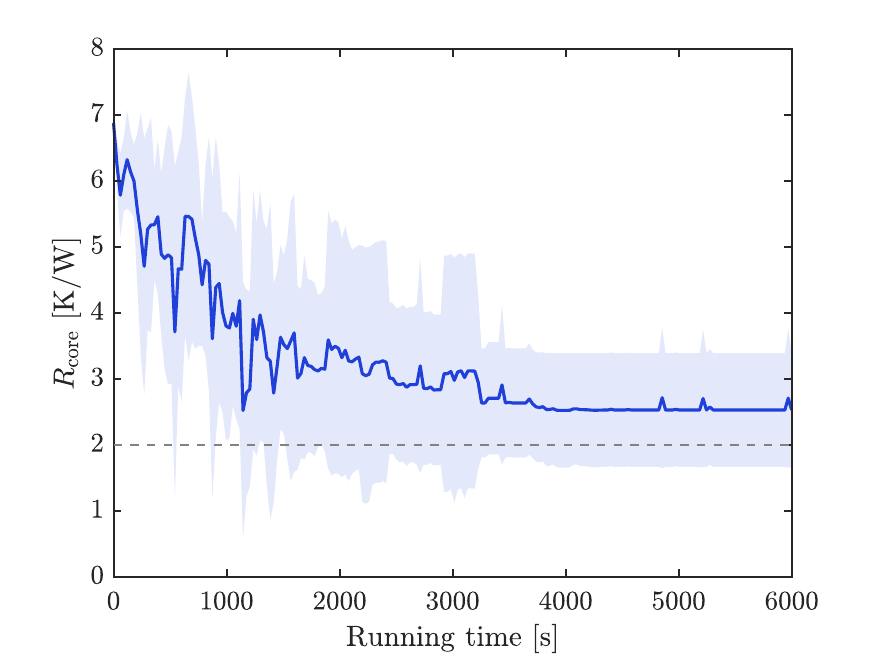}}
    \subfloat{
    \centering
    \includegraphics[width = .24\textwidth,trim={.2cm 0cm .9cm .3cm},clip]{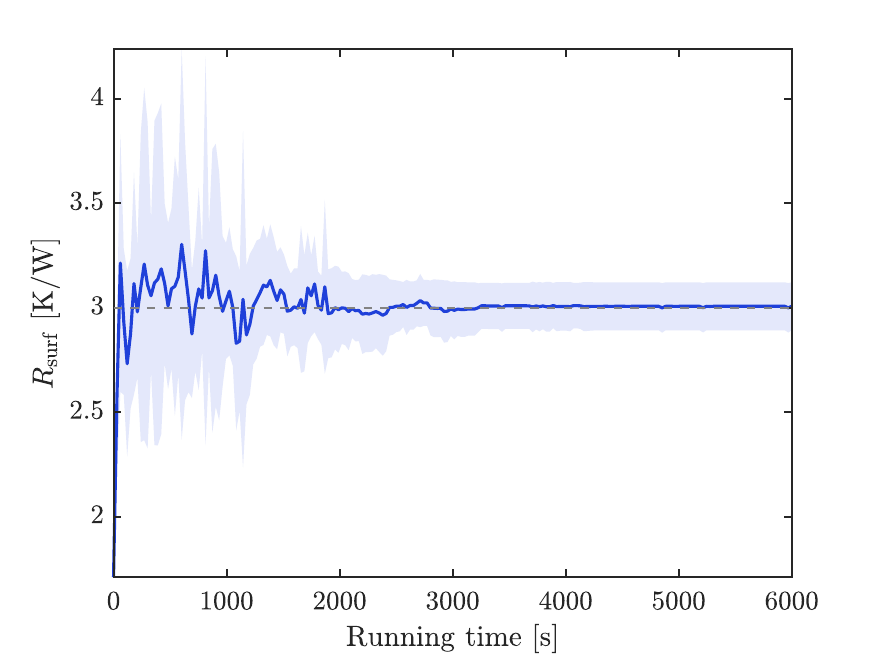}}

     \subfloat{
    \centering
    \includegraphics[width = .24\textwidth,trim={.2cm 0cm .9cm .3cm},clip]{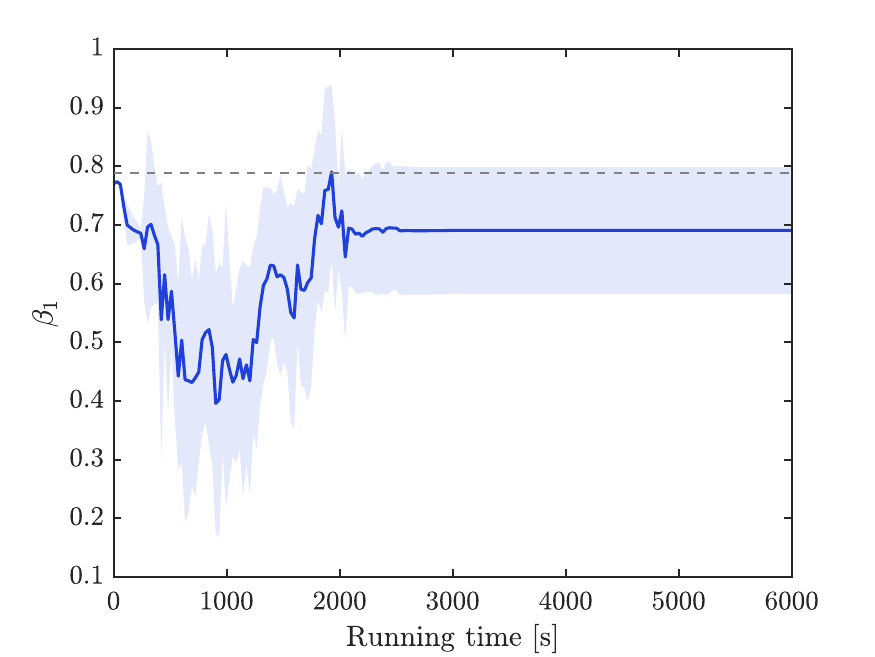}}
    \subfloat{
    \centering
    \includegraphics[width = .24\textwidth,trim={.2cm 0cm .9cm .3cm},clip]{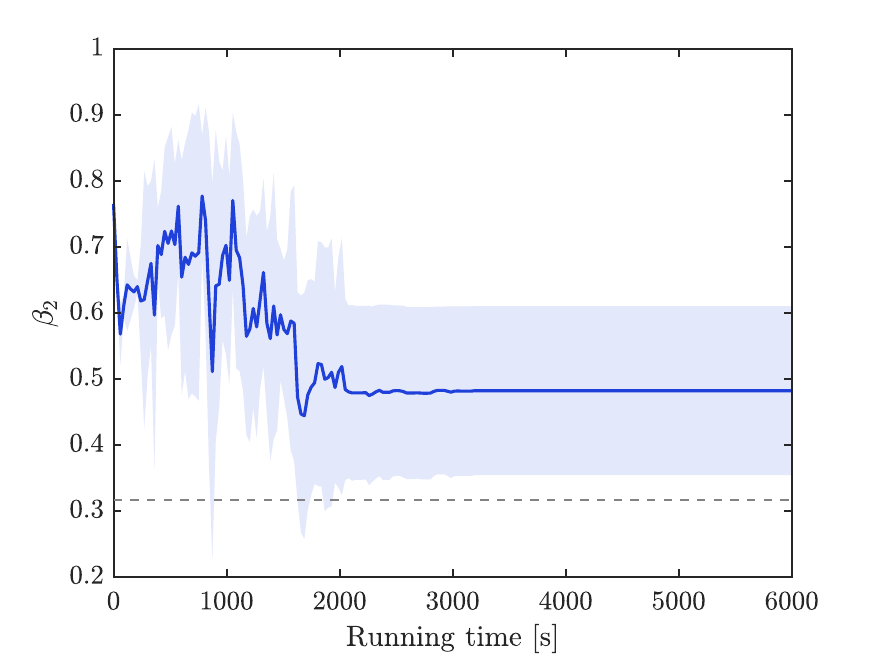}}
    \subfloat{
    \centering
    \includegraphics[width = .24\textwidth,trim={.2cm 0cm .9cm .3cm},clip]{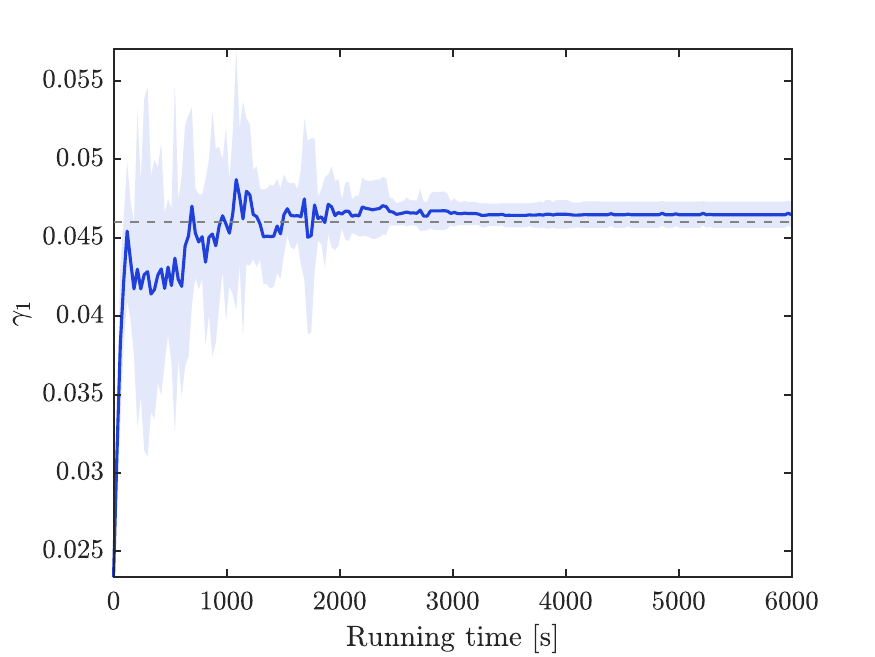}}
     \subfloat{
    \centering
    \includegraphics[width = .24\textwidth,trim={.2cm 0cm .9cm .3cm},clip]{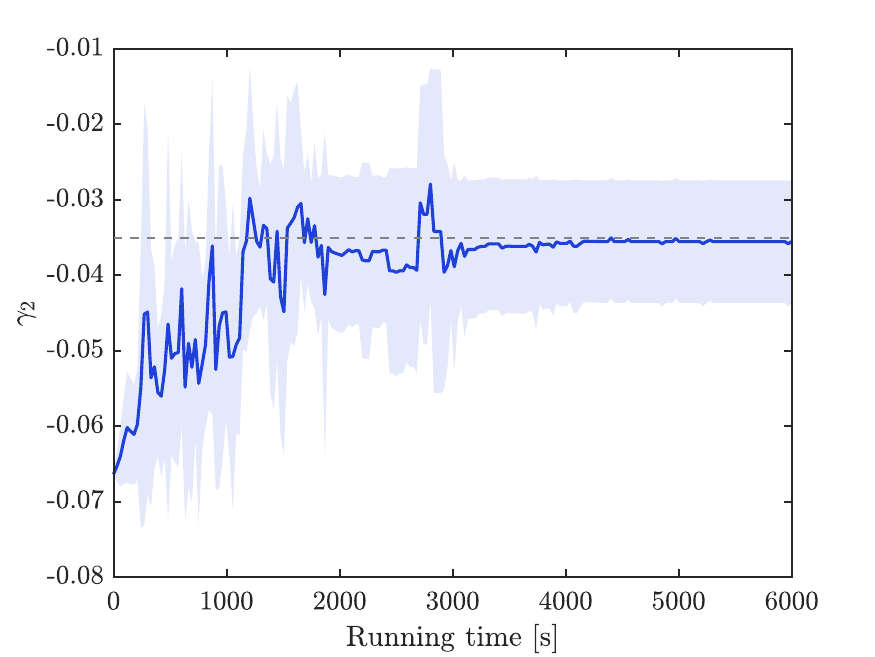}}

    \subfloat{
    \centering
    \includegraphics[width = .24\textwidth,trim={.2cm 0cm .9cm .3cm},clip]{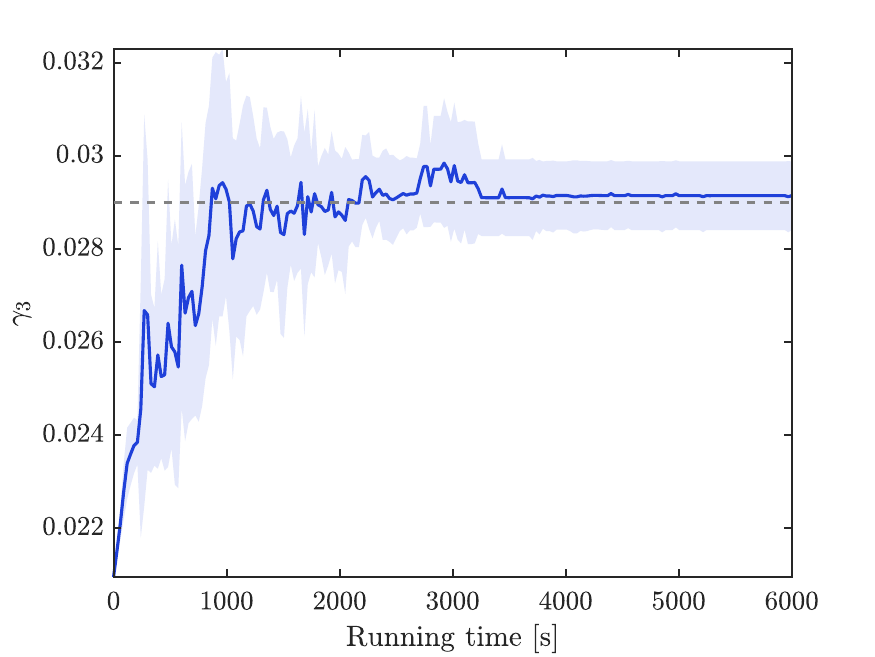}}
    \subfloat{
    \centering
    \includegraphics[width = .24\textwidth,trim={.2cm 0cm .9cm .3cm},clip]{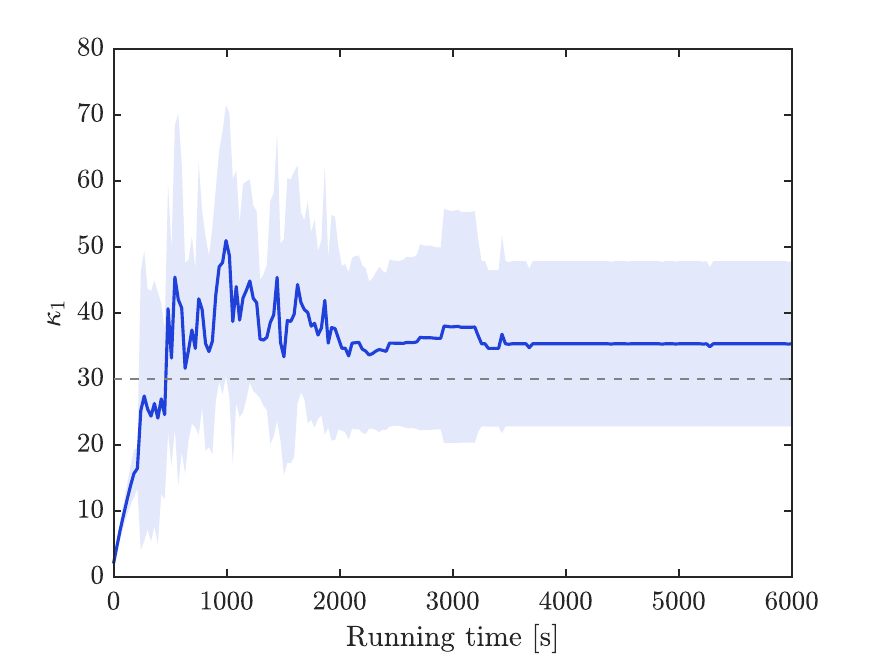}}
    \subfloat{
    \centering
    \includegraphics[width = .24\textwidth,trim={.2cm 0cm .9cm .3cm},clip]{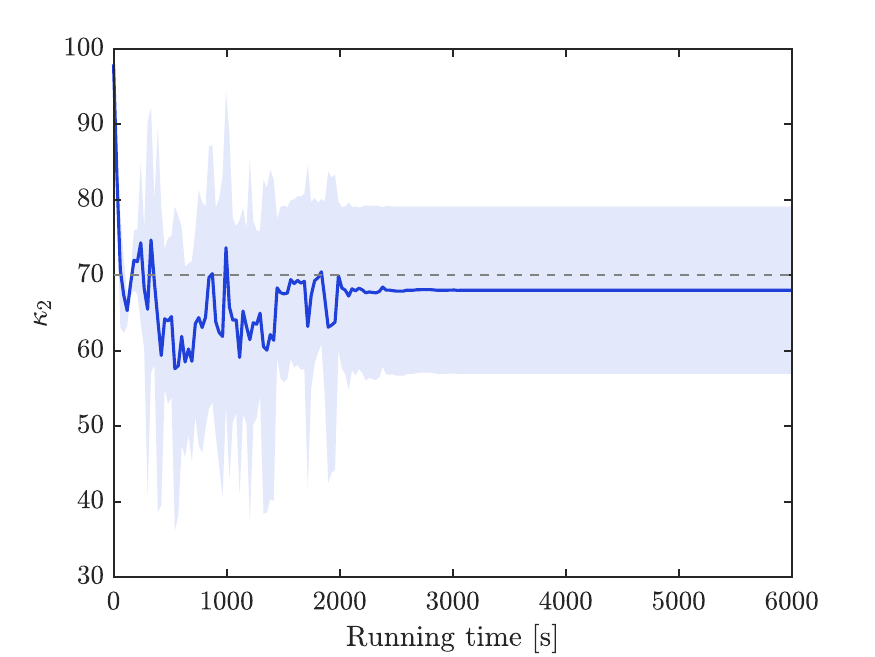}}
    \subfloat{
    \centering
    \includegraphics[width = .24\textwidth,trim={.2cm 0cm .9cm .3cm},clip]{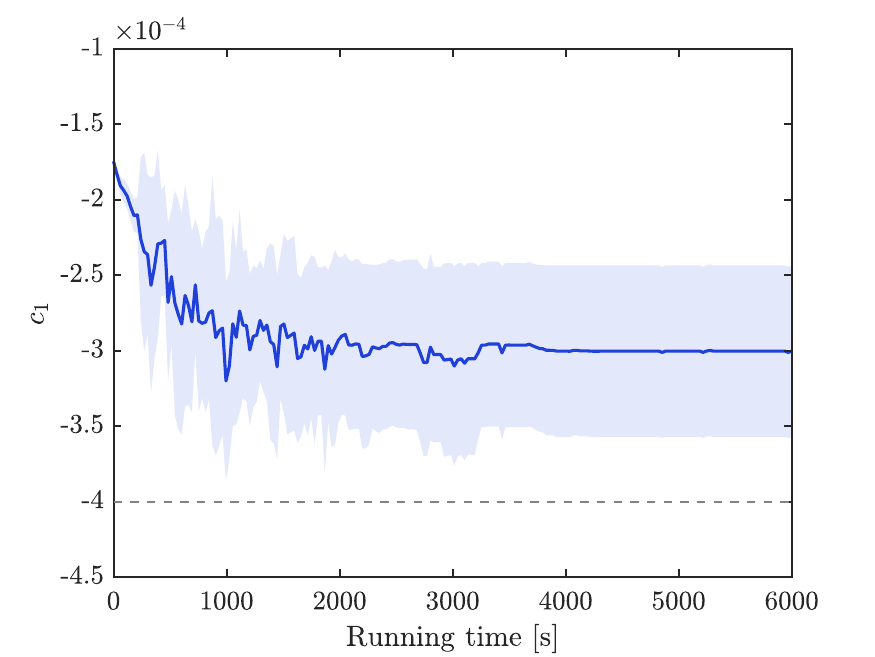}}

    \subfloat{
    \centering
    \includegraphics[width = .24\textwidth,trim={.2cm 0cm .9cm .3cm},clip]{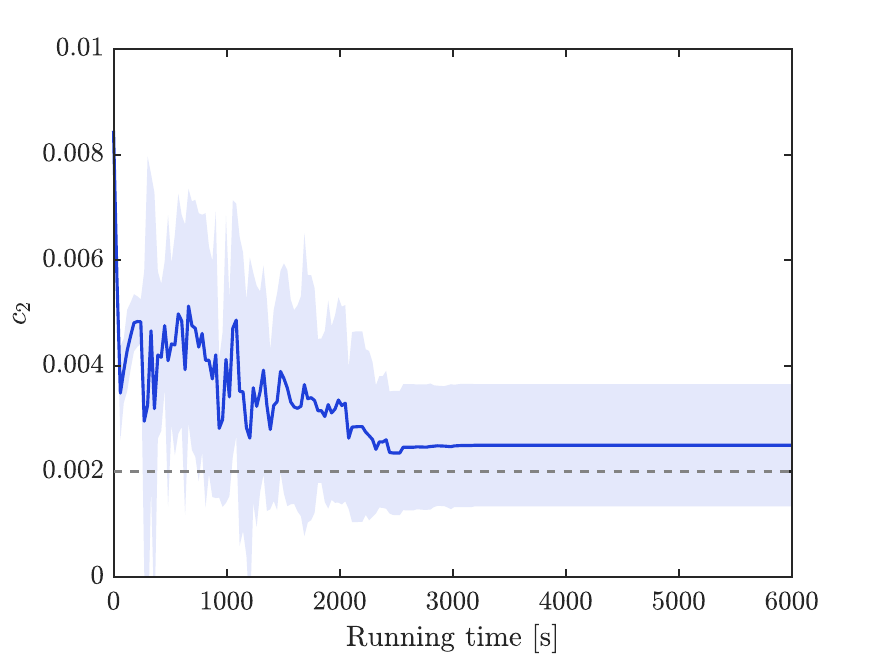}}
    \subfloat{
    \centering
    \includegraphics[width = .24\textwidth,trim={.2cm 0cm .9cm .3cm},clip]{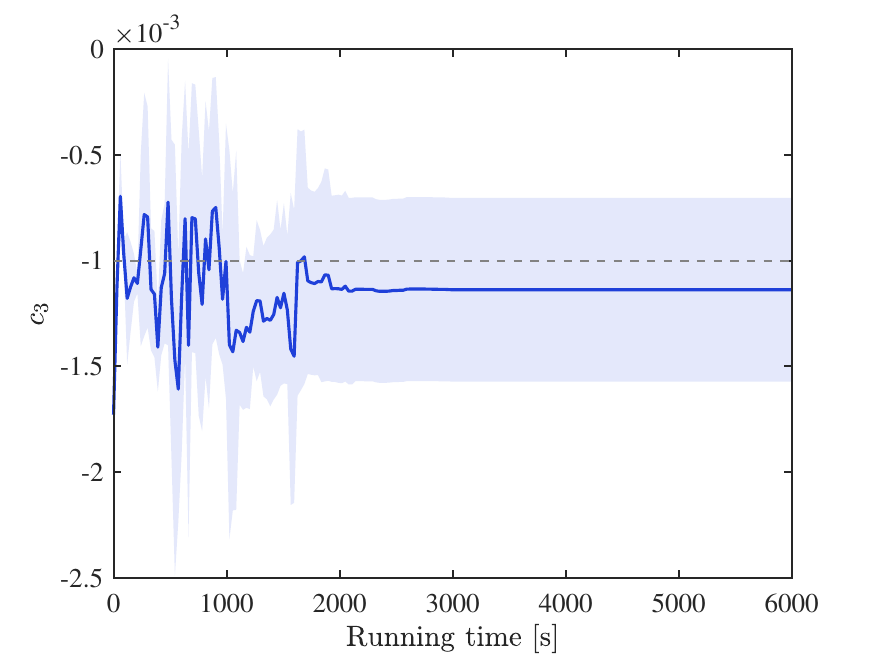}}
    
    \caption{Convergence of the 18 parameters across 50 numerical simulation runs. The dashed line indicates the nominal parameter value $\param^*$ used to generate the measurement data.}
    \label{Fig: Box plot}
\end{figure*}

This simulation study uses the BattX model identified in~\cite{Biju:AE:2023} for a Samsung INR18650-25R cell with NCA cathode and graphite anode as the nominal model (Table~\ref{Table: Sim result}). Synthetic datasets are generated from this model, and the proposed accelerated BayesOpt approach is then applied to recover the BattX model, with a comparison with the nominal model. The simulation setting is as follows.

\begin{itemize}

\item The synthetic datasets for model identification are generated by running the nominal model under the constant-current discharging profiles at 3 C ($T_\mathrm{amb} = 303\, \si{K}$), 4 C ($T_\mathrm{amb} = 293\, \si{K}$), 5 C ($T_\mathrm{amb} = 283\, \si{K}$),  as well as variable-current profiles including UDDS ($T_\mathrm{amb} = 283\, \si{K}$), US06 ($T_\mathrm{amb} = 293\, \si{K}$), and SC04 ($T_\mathrm{amb} = 303\, \si{K}$).

\item To validate the identified model, additional synthetic datasets are generated by running the nominal model with a variable-current LA92 profile and a notional electric vertical takeoff and landing (eVTOL) operation profile, with $T_\mathrm{amb} = 298\, \si{K}$ for both cases. 

\item All variable-current profiles are scaled to be between $0\sim5$ C, and for all the datasets, the initial SoC is $\mathrm{SoC}(0) = 100\%$. 

\item The noise covariances are $\Qc = 10^{-8}\mathbf{I}$ and $\Rk = \mathrm{diag}(10^{-3},10^{-2})$. The sampling interval is set to be 1 $\mathrm{s}$.

\item All computations are performed in MATLAB on a host computer equipped with a 3.2 GHz Intel\textsuperscript{\textregistered} i9-12900KF CPU and 128 GB RAM.

\end{itemize}

For evaluation, we apply the accelerated BayesOpt, standard BayesOpt, and Nelder--Mead methods to extract the BattX model from the synthetic datasets.  All three methods use the U-IPF method with $N_p = 100$ particles for likelihood evaluation, and each is executed for 100 independent runs.  Fig.~\ref{Fig:AlgComp} shows the evaluation of the average $L(\param)$ across the 100 runs, together with the approximate standard-deviation bounds, from which several observations arise. 

\begin{itemize}

\item The accelerated BayesOpt approach consistently outperforms the other methods. It almost always attains the highest likelihood and exhibits a rapid increase in $L(\param)$, indicating both fast convergence and strong optimization performance.

\item In comparison, the standard BayesOpt approach  converges more slowly and does not always reach the highest likelihood within the allotted time, despite its theoretical potential for global optimization. This behavior reflects the practical challenges faced by BayesOpt in navigating high-dimensional parameter spaces.

\item The Nelder--Mead method demonstrates fast convergence but typically converges to local optima. While aligning with its design, this tendency highlights the method's limitation when global optimality is essential for accurate and physically consistent BattX model identification. 

\item For additional comparison, we also apply the particle swarm optimization method~\cite{Yu:TIE:2017} and a direct gradient-based method~\cite{Poyiadjis:ACC:2006}. As shown in Fig.~\ref{Fig:AlgComp_b}, Both exhibit relatively slow convergence and, in most cases, fail to reach the global optimum within the given time.

\end{itemize}

These results underscore the advantages of the accelerated BayesOpt approach, which combines fast convergence and low computational overhead with reliable attainment of near-global optima in practice. For further assessment, Table~\ref{Table: Sim result} reports the parameter estimates obtained by the accelerated BayesOpt approach alongside the nominal values, and Fig.~\ref{Fig: Box plot} presents the convergence in the estimation of each parameter. The estimated parameters are generally close to their nominal values, although the estimation accuracy varies across the parameters and certain biases exist for some parameters, due to differences in the identifiability of individual parameters. 

In this study, we employ the U-IPF method to evaluate the likelihood function. To assess its effectiveness and practical utility, we compare it with the APF. To this end, we run each method 1,000 times on the nominal BattX model using $N_p = 10$, $100$, and $1000$, respectively. Fig.~\ref{Fig:Likelihood comp} presents the boxplots of the resulting likelihood estimates. The results clearly indicate that, in all cases, U-IPF provides significantly more accurate likelihood evaluations with substantially smaller variance. For both methods, increasing the number of particles improves accuracy; however, the performance of U-IPF with only $N_p=10$ particles is already markedly better than that of the APF with $N_p=1,000$ particles. This level of accuracy ensures  more effective parameter search, while the ability to operate with far fewer particles greatly reduces the computational overhead.

\subsection{Experimental Validation}

\begin{figure}[t!]
    \centering
    \subfloat[]{
    \centering
    \includegraphics[width = .35\textwidth,trim={0cm 3cm 0cm 7cm},clip]{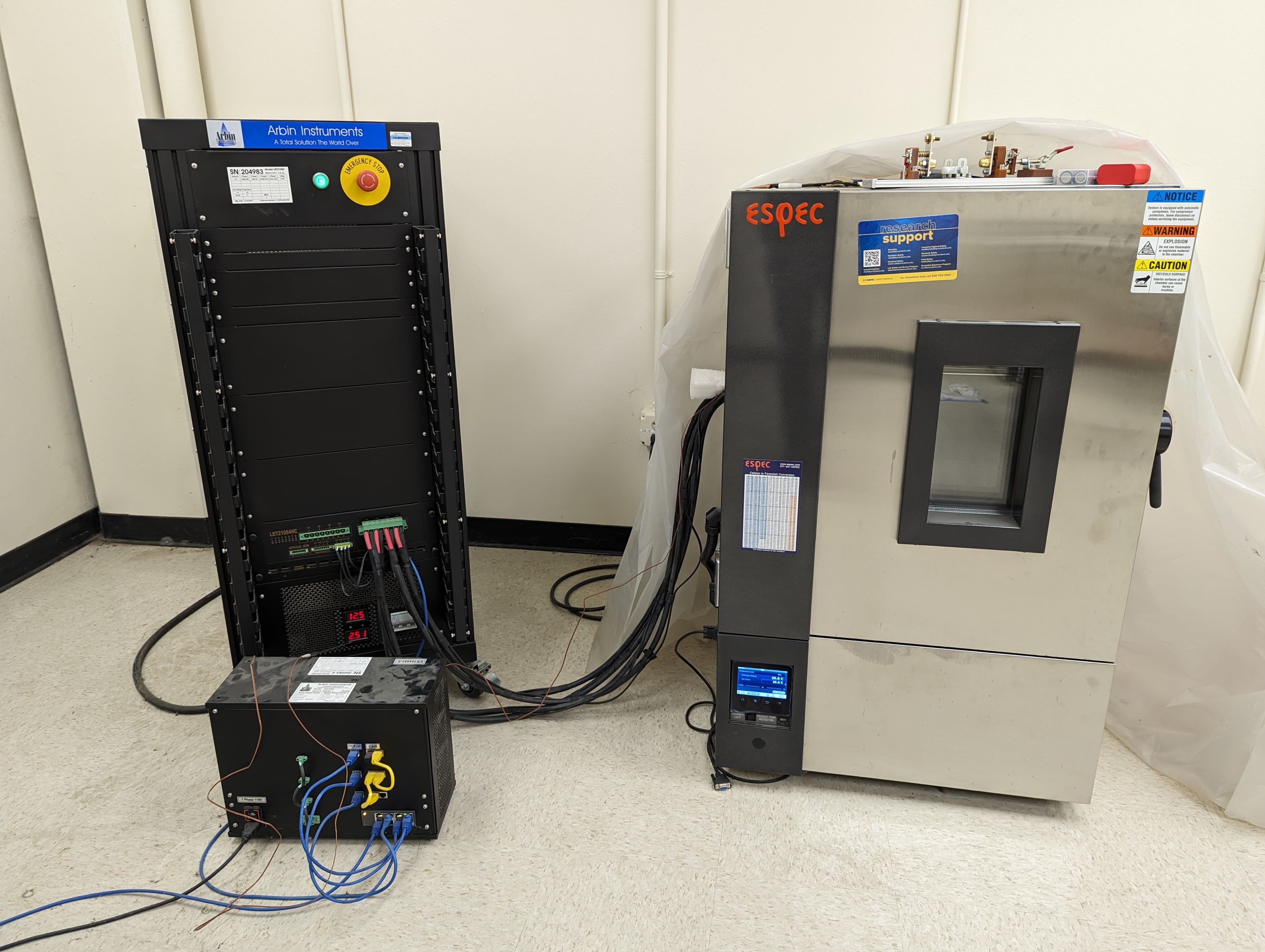}}
    
    \subfloat[]{
    \centering
    \includegraphics[width = .35\textwidth,trim={0cm 5cm 0cm 5cm},clip]{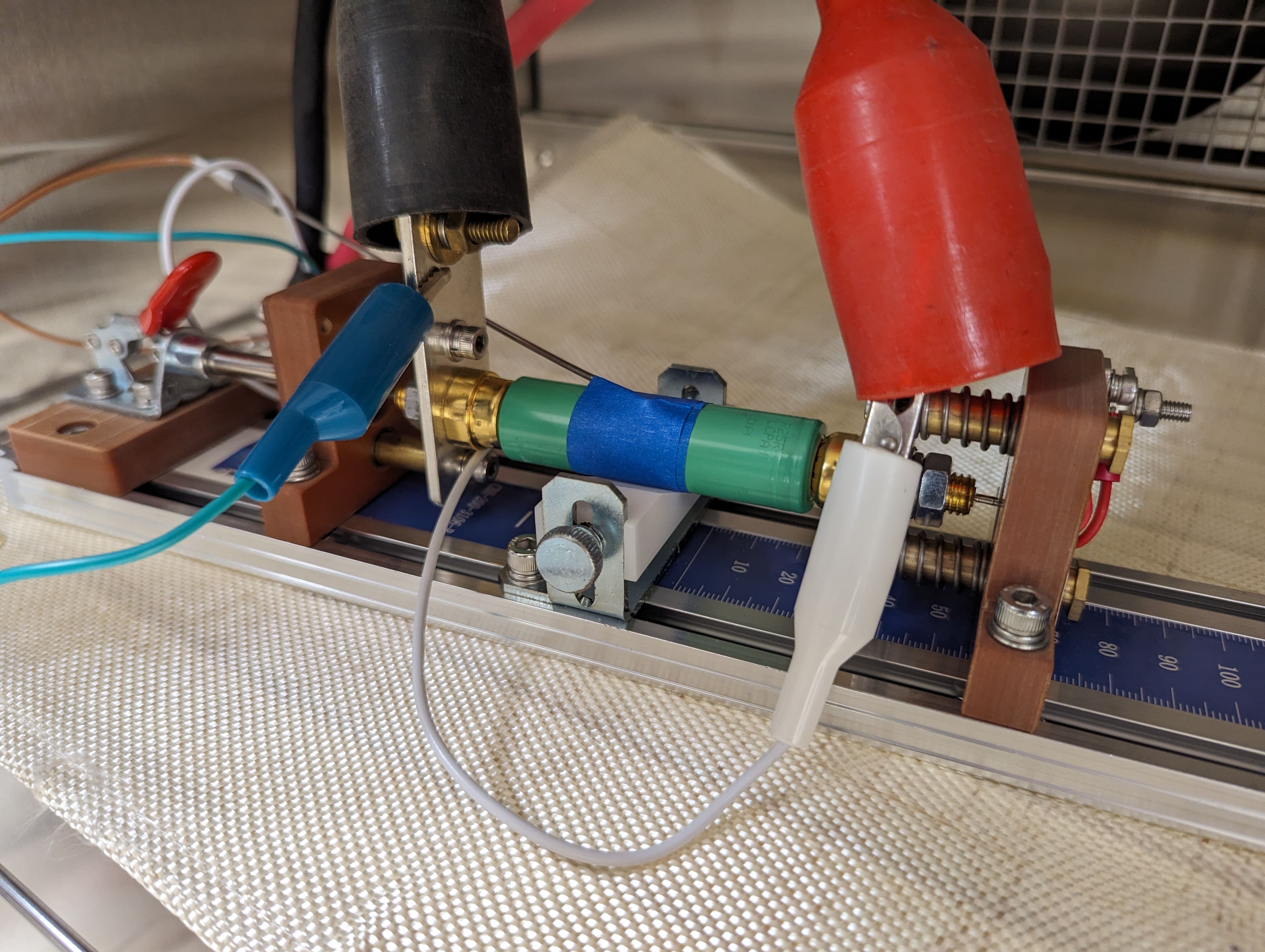}}
    \caption{Experimental setup for data collection: (a) the Arbin battery tester and the ESPEC thermal chamber; (b) the Samsung INR18650-25R cell under testing.}
    \label{Fig: Arbin tester}
\end{figure}

\begin{table}[t!]\centering
\ra{1.2}
\caption{Parameter identification results in the experimental validation.}
 \begin{tabular}{ l | l l l }
\toprule
Parameters & Search range &  Identified \\
\midrule
$C_{s,1}\ [\si{F}]$ & 3000$\sim$8000 & 6423.2624  \\
$R_{s,1}\ [\si{\Omega}]$ & 0$\sim$1 & 0.6479 \\
$C_{e}\ [\si{F}]$ & 0$\sim$8000 & 4567.8173 \\
$R_{e}\ [\si{\Omega}]$ & 0$\sim$0.1 & 0.0142 \\
$C_{\mathrm{core}}\ [\si{J/K}]$ & 0$\sim$100 & 59.2864 \\
$C_{\mathrm{surf}}\ [\si{J/K}]$ & 0$\sim$50 & 16.6883 \\
$R_{\mathrm{core}}\ [\si{K/W}]$ & 0$\sim$10 & 0.7369 \\
$R_{\mathrm{surf}}\ [\si{K/W}]$ & 0$\sim$10 & 2.9218 \\
$\beta_1$ & 0$\sim$200 & 55.9396 \\
$\beta_2$ & 0$\sim$200 & 133.2875 \\
$\gamma_1\ $ & 0$\sim$0.1 & 0.0529 \\
$\gamma_2\ $ & -0.1$\sim$0 & -0.0311 \\
$\gamma_3\ $ & 0$\sim$0.1 & 0.0266 \\
$\kappa_1$ & 0$\sim$10000 & 2414.5960 \\
$\kappa_2$ & 0$\sim$10000 & 5085.8376 \\
$c_1\ $ & -0.001$\sim$0 & -0.0004 \\
$c_2\ $ & 0$\sim$0.01 & 0.0017 \\
$c_3\ $ & -0.01$\sim$0 & -0.0012 \\
\bottomrule
\end{tabular}
\label{Table: Exp result}
\end{table}

\begin{table}[t!]\centering
\ra{1.2}
\caption{Prediction accuracy of the identified BattX model on two experimental validation datasets.}
 \begin{tabular}{ l c c }
\toprule
 \makecell{Current \\ profile} & \makecell{ Voltage \\ RMSE } & \makecell{Temperature \\ RMSE} \\
\midrule
 LA92 & 17.25 mV & 0.15 $\mathrm{K}$ \\
 eVTOL & 14.93 mV & 0.15 $\mathrm{K}$ \\
\bottomrule
\end{tabular}

\label{Table: RMSE}
\end{table}

\begin{figure}[t!]
    \centering
    \includegraphics[width = 0.48\textwidth,trim={0cm 0cm 0cm 0cm},clip]{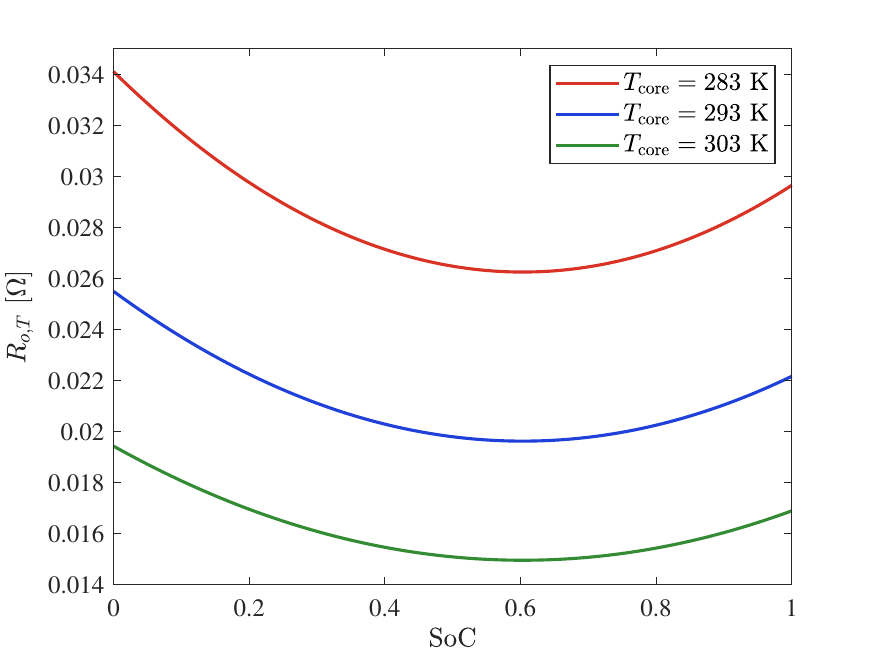}
    \caption{Identified $R_{o,T}$ at various SoC values and core temperatures.}
    \label{Fig:R0}
\end{figure}

Proceeding forward, we conduct an experimental validation by applying the accelerated BayesOpt approach to experimental data for BattX model identification. The experimental setup is shown in Fig.~\ref{Fig: Arbin tester}. Specifically, a Samsung INR18650-25R cell with NCA cathode and graphite anode is tested using an Arbin LBT21084 battery tester in conjunction with an ESPEC thermal chamber. A thermocouple attached to the surface of the cell is used to measure its temperature during the experiments. We collect the experimental datasets for both model identification and validation, using the same current profiles and ambient temperatures as those employed in the simulation study in Section~\ref{sec: numerical simulation}.

\begin{figure*}[t!]
\centering
    \subfloat[]{
    \centering
    \includegraphics[width = .48\textwidth,trim={1cm .7cm 1cm .9cm},clip]{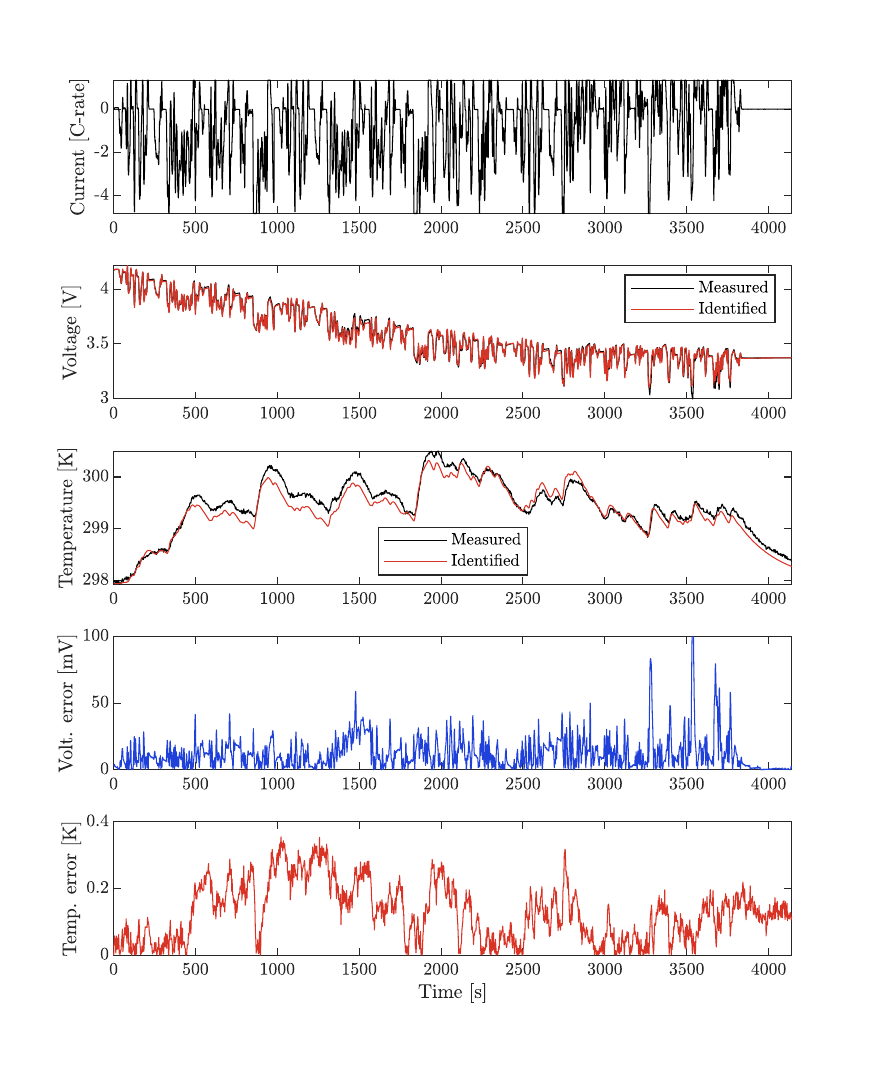}}
    \subfloat[]{
    \centering
    \includegraphics[width = .48\textwidth,trim={1cm .7cm 1cm .9cm},clip]{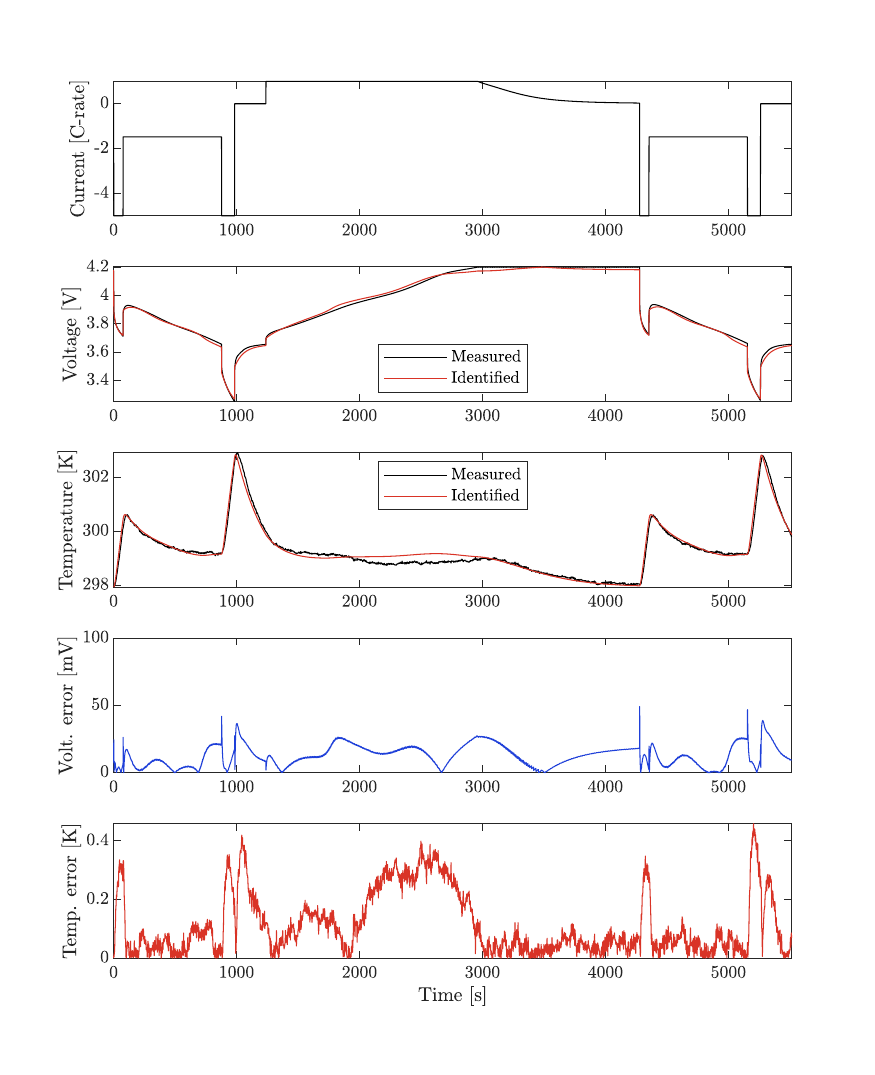}}
    \caption{Voltage and temperature prediction by the identified BattX model versus measurements for (a) the LA92 profile and (b) the notional eVTOL profile at $T_{\mathrm{amb}} = 298~\si{K}$.}
    \label{Fig: Exp results}
\end{figure*}

Given the experimental datasets, the accelerated BayesOpt approach is executed together with the U-IPF-based likelihood evaluation, using the parameter search ranges listed in Table~\ref{Table: Exp result}. The identification results are summarized as follows. First, Table~\ref{Table: Exp result} reports the identified parameter values, and Fig.~\ref{Fig:R0} illustrates the estimated internal resistance  $R_{o,T}$, as well as its dependence on temperature and SoC. These estimation results agree with established understanding of the cell behavior~\cite{Biju:AE:2023}. We further assess the predictive capability of the identified BattX model by comparing its predictions against independent validation datasets obtained under the LA92 current profile and an eVTOL operation profile. Fig.~\ref{Fig: Exp results} shows the predicted voltage and temperature versus the corresponding measurements. For both datasets, the predictions closely match the measurements, as further corroborated by the prediction errors shown in Fig.~\ref{Fig: Exp results}. Table~\ref{Table: RMSE} reports the root-mean-square errors (RMSEs) for voltage and temperature prediction. The low RMSE values in both cases substantiate the effectiveness of the accelerated BayesOpt approach for accurate model identification. It is worth noting that the identified parameter values differ slightly from those reported in~\cite{Biju:AE:2023}. This discrepancy is likely due to the use of datasets collected under a broader range of test profiles and ambient temperatures in the present study.

\section{Conclusions} \label{sec: Conclusion}

SSMs are foundational for complex system representation, analysis, monitoring, and control. However, their parameter identification is often challenging due to nonlinearity, high dimensionality, and the lack of analytical gradients.   BayesOpt   has recently emerged as a promising approach for addressing SSM identification, owing to its derivative-free nature and global search capability. Nevertheless, its practical deployment is often hindered by slow convergence and high computational cost. To address these limitations, this paper presents a new framework that enhances BayesOpt by integrating it with the Nelder–Mead method. This framework exploits their complementary strengths: BayesOpt provides global, uncertainty-aware exploration, while the Nelder–Mead method offers fast and computationally efficient local search. We develop principled initialization, switching and termination rules to coordinate the two methods within a unified optimization framework. The resulting hybrid approach accelerates convergence, mitigates computational overhead, and preserves global search capability. To further enhance efficiency, we incorporate the U-IPF method for accurate and fast likelihood evaluation within the proposed framework. We validate the proposed approach by applying it to the identification of the BattX  model for LiBs, which includes 18 unknown parameters. Both simulation and experimental results demonstrate its effectiveness in extracting the model parameters from data.

\appendix
\renewcommand{\theequation}{A.\arabic{equation}}

Here, we provide a brief review of the BattX model and refer the reader to~\cite{Biju:AE:2023} for further details. The BattX model is designed to capture the behavior of a LiB cell operating across low to high C-rates. It couples four circuits, each emulating a major physical process during charging and discharging.

Sub-circuit A simulates the lithium-ion diffusion process within the electrode. The capacitors $C_{s,i}$ for $i=1,\ldots,N$ store charge, with their total capacitance equal to the cell capacity. Charge transfer among these capacitors mimics lithium-ion diffusion within the electrode, while the resistors $R_{s,i,T}$ for $i=1,\ldots,N$ represent the resistance to this transfer, similar to the opposition to the transport of lithium ions in the electrode. The voltages $V_{s,i}$ across $C_{s,i}$ for $i=1,\ldots,N$ correspond to the spatial distribution of lithium-ion concentration within the electrode. Their governing equations are
\begin{align*}
    \dot{V}_{s,1} &= \frac{{V}_{s,2}-{V}_{s,1}}{C_{s,1}R_{s,1,T}} + \frac{I}{C_{s,1}}, \\
    \dot{V}_{s,i} &= \frac{{V}_{s,i-1}-{V}_{s,i}}{C_{s,i}R_{s,i-1,T}} + \frac{{V}_{s,i+1}-{V}_{s,i}}{C_{s,i}R_{s,i,T}},\ i=2,\dots,N-1 \\
    \dot{V}_{s,N} &= \frac{{V}_{s,N-1}-{V}_{s,N}}{C_{s,N}R_{s,N,T}},
\end{align*}
where $I$ is the input current with $I<0$ for discharging and $I>0$ for charging. By limiting $0\le{V}_{s,i}\le1$ for normalization, the SoC can be defined as
\begin{align*}
    \mathrm{SoC} = \frac{\sum_{i=1}^N C_{s,i}V_{s,i}}{\sum_{i=1}^N C_{s,i}} \times 100\%.
\end{align*}  

Sub-circuit B employs another resistor-capacitor chain to emulate lithium-ion diffusion within the electrolyte, following a modeling approach similar to that of sub-circuit A. Its dynamics is governed by
\begin{align*}
    \dot{V}_{e,1} &= \frac{{V}_{e,2}-{V}_{e,1}}{C_{e}R_{e}} + \frac{I}{C_{e}}, \\
    \dot{V}_{e,2} &= \frac{{V}_{e,1}-2{V}_{e,2}+{V}_{e,3}}{C_{e}R_{e}}, \\
    \dot{V}_{e,3} &= \frac{{V}_{e,2}-{V}_{e,3}}{C_{e}R_{e}} - \frac{I}{C_{e}}.
\end{align*}

Sub-circuit C is a lumped thermal model for capturing the temperature dynamics of the cell. It lumps the spatial temperature distribution into two representative points, namely the cell core and the surface. The corresponding temperatures,  $T_\mathrm{core}$ and $T_\mathrm{surf}$, obey the following dynamics:
\begin{align*}
    \dot{T}_\mathrm{core} &= \frac{\dot{Q}}{C_\mathrm{core}} + \frac{T_\mathrm{surf}-T_\mathrm{core}}{R_\mathrm{core}C_\mathrm{core}}, \\
    \dot{T}_\mathrm{surf} &= \frac{T_\mathrm{amb}-T_\mathrm{surf}}{R_\mathrm{surf}C_\mathrm{surf}} - \frac{T_\mathrm{surf}-T_\mathrm{core}}{R_\mathrm{core}C_\mathrm{surf}}, 
\end{align*}
where $C_\mathrm{core}$ and $C_\mathrm{surf}$ are the thermal capacitances,  $R_\mathrm{core}$ and $R_\mathrm{surf}$ are the thermal resistances, and $T_\mathrm{amb}$ is the ambient temperature. The heat generation rate $\dot{Q}$ is described as 
\begin{align*}
    \dot{Q} = I\left(V-U_{s}(\mathrm{SoC})\right) + I T_\mathrm{core}\frac{dU_{s}}{dT_\mathrm{core}},
\end{align*}
where $V$ is the cell's terminal voltage, $U_{s}$ is the OCV function. In the above equation, the first term represents irreversible ohmic heating, and the second term represents reversible entropic heating with
\begin{align*}
    \frac{dU_{s}}{dT_\mathrm{core}} = c_1 +c_2\cdot \mathrm{SoC} +c_3 \cdot \mathrm{SoC}^2 ,
\end{align*}
where $c_1$,$c_2$,$c_3$ are coefficients. 

Finally, sub-circuit D models the LiB cell's terminal voltage:
\begin{align*}
    V = U_{s}(V_{s,1}) + U_e(V_{e,1}, V_{e,3}) + R_{o,T} I.
\end{align*}
The first term represents the OCV due to electrode dynamics, the second term the voltage contribution from electrolyte dynamics, and the third term the voltage drop across the internal resistance $R_{o,T}$. Here, the form of $U_e$ is designed based on the SPMeT as
\begin{align*}
    U_{e} = \beta_1 \left( \ln\left(\frac{V_{e,1}+\beta_2}{V_{e,3}+\beta_2}\right)\right).
\end{align*}
Further, $R_{o,T}$ is both SoC- and temperature-dependent, following
\begin{align*}
    R_{o,T} = R_o(\mathrm{SoC}) \cdot \exp\left(\kappa_1 \left( \frac{1}{T_\mathrm{core}} - \frac{1}{T_\mathrm{ref}} \right) \right),
\end{align*}
where $T_\mathrm{ref}$ is the reference temperature, and $R_o(\mathrm{SoC})$ is given by
\begin{align*}
    R_o(\mathrm{SoC}) = \gamma_1 + \gamma_2\cdot\mathrm{SoC} + \gamma_3\cdot\mathrm{SoC}^2.
\end{align*}
Since temperature influences diffusion, $R_{s,i,T}$ for $i=1,\ldots,N-1$ are also modeled as temperature dependent:
\begin{align*}
    R_{s,i,T} = R_{s,i} \cdot \exp\left(\kappa_2 \left( \frac{1}{T_\mathrm{core}} - \frac{1}{T_\mathrm{ref}} \right) \right)
\end{align*}

Together, the above equations provide a complete SSM representation of the BattX model used in Section~\ref{sec: LiB identification}.

\bibliographystyle{ieeetr}
\bibliography{ref}

@article{Cheng:CBC:2015,
title = {Ancestral population genomics using coalescence hidden {M}arkov models and heuristic optimisation algorithms},
journal = {Computational Biology and Chemistry},
volume = {57},
pages = {80-92},
year = {2015},
issn = {1476-9271},
author = {Jade Yu Cheng and Thomas Mailund}
}

@INPROCEEDINGS{Pi:MECC24:2024,
title = {Parameter Identification for Electrochemical Models of Lithium-Ion Batteries Using {B}ayesian Optimization},
pages = {180-185},
year = {2024},
booktitle = {The 4th Modeling, Estimation, and Control Conference},
author = {Jianzong Pi and Samuel Filgueira {da Silva} and Mehmet Fatih Ozkan and Abhishek Gupta and Marcello Canova},
}

@article{Nelder:TCJ:1965,
    author = {Nelder, J. A. and Mead, R.},
    title = {A Simplex Method for Function Minimization},
    journal = {The Computer Journal},
    volume = {7},
    number = {4},
    pages = {308-313},
    year = {1965},
    month = {01},
    issn = {0010-4620},
    doi = {10.1093/comjnl/7.4.308},
    url = {https://doi.org/10.1093/comjnl/7.4.308},
    eprint = {https://academic.oup.com/comjnl/article-pdf/7/4/308/1013182/7-4-308.pdf},
}

@ARTICLE{Ljung:TAC:1979,
  author={Ljung, L.},
  journal={IEEE Transactions on Automatic Control}, 
  title={Asymptotic behavior of the extended {K}alman filter as a parameter estimator for linear systems}, 
  year={1979},
  volume={24},
  number={1},
  pages={36-50},
  doi={10.1109/TAC.1979.1101943}}

@ARTICLE{Cox:TAC:1964,
  author={Cox, H.},
  journal={IEEE Transactions on Automatic Control}, 
  title={On the estimation of state variables and parameters for noisy dynamic systems}, 
  year={1964},
  volume={9},
  number={1},
  pages={5-12},
  doi={10.1109/TAC.1964.1105635}}

@article{Kalman:IFAC:1960,
title = {On the general theory of control systems},
journal = {IFAC Proceedings Volumes},
volume = {1},
number = {1},
pages = {491-502},
year = {1960},
note = {1st International IFAC Congress on Automatic and Remote Control, Moscow, USSR, 1960},
issn = {1474-6670},
doi = {https://doi.org/10.1016/S1474-6670(17)70094-8},
author = {R.E. Kalman}
}

@article{frazier:tutorial:2018,
title={A Tutorial on {B}ayesian Optimization}, 
author={Peter I. Frazier},
journal={arXiv:1807.02811},
year={2018}
}

@article{ASKARI:AUTO:2022,
title = {Implicit particle filtering via a bank of nonlinear {K}alman filters},
journal = {Automatica},
volume = {145},
pages = {110469},
year = {2022},
issn = {0005-1098},
doi = {https://doi.org/10.1016/j.automatica.2022.110469},
author = {Iman Askari and Mulugeta A. Haile and Xuemin Tu and Huazhen Fang}
}

@article{KantasL:SS:2015,
author = {Nikolas Kantas and Arnaud Doucet and Sumeetpal S. Singh and Jan Maciejowski and Nicolas Chopin},
title = {On Particle Methods for Parameter Estimation in State-Space Models},
volume = {30},
journal = {Statistical Science},
number = {3},
publisher = {Institute of Mathematical Statistics},
pages = {328 -- 351},
year = {2015},
doi = {10.1214/14-STS511},
}

@article{Chorin:Math:2010,
author = {Alexandre Chorin and Matthias Morzfeld and Xuemin Tu},
title = {Implicit particle filters for data assimilation},
volume = {5},
journal = {Communications in Applied Mathematics and Computational Science},
number = {2},
publisher = {MSP},
pages = {221 -- 240},
year = {2010},
}

@INPROCEEDINGS{Tu:ACC:2024,
  author={Tu, Hao and Lin, Xinfan and Wang, Yebin and Fang, Huazhen},
  booktitle={American Control Conference}, 
  title={System Identification for Lithium-Ion Batteries with Nonlinear Coupled Electro-Thermal Dynamics via {B}ayesian Optimization}, 
  year={2024},
  volume={},
  number={},
  pages={1946-1951},
  doi={10.23919/ACC60939.2024.10645049}
}

@article{Biju:AE:2023,
title = {Batt{X}: An equivalent circuit model for lithium-ion batteries over broad current ranges},
journal = {Applied Energy},
volume = {339},
pages = {120905},
year = {2023},
issn = {0306-2619},
doi = {https://doi.org/10.1016/j.apenergy.2023.120905},
url = {https://www.sciencedirect.com/science/article/pii/S0306261923002696},
author = {Nikhil Biju and Huazhen Fang}
}

@inproceedings{NEURIPS:David:2019,
 author = {Eriksson, David and Pearce, Michael and Gardner, Jacob and Turner, Ryan D and Poloczek, Matthias},
 booktitle = {Advances in Neural Information Processing Systems},
 pages = {},
 publisher = {},
 title = {Scalable Global Optimization via Local {B}ayesian Optimization},
 volume = {32},
 year = {2019}
}

@article{Lin:JPS:2014,
title = {A lumped-parameter electro-thermal model for cylindrical batteries},
journal = {Journal of Power Sources},
volume = {257},
pages = {1-11},
year = {2014},
issn = {0378-7753},
doi = {https://doi.org/10.1016/j.jpowsour.2014.01.097},
author = {Xinfan Lin and Hector E. Perez and Shankar Mohan and Jason B. Siegel and Anna G. Stefanopoulou and Yi Ding and Matthew P. Castanier}
}

@article{Schon:Auto:2011,
title = {System identification of nonlinear state-space models},
journal = {Automatica},
volume = {47},
number = {1},
pages = {39-49},
year = {2011},
issn = {0005-1098},
doi = {https://doi.org/10.1016/j.automatica.2010.10.013},
url = {https://www.sciencedirect.com/science/article/pii/S0005109810004279},
author = {Thomas B. Schön and Adrian Wills and Brett Ninness}
}

@article{Dahlin:arXiv:2017,
      title={{B}ayesian optimisation for fast approximate inference in state-space models with intractable likelihoods}, 
      author={Johan Dahlin and Mattias Villani and Thomas B. Schön},
      year={2017},
      eprint={1506.06975},
      archivePrefix={arXiv},
      primaryClass={stat.CO}
}

@ARTICLE{Mahdi:TNNLS:2022,
  author={Imani, Mahdi and Ghoreishi, Seyede Fatemeh},
  journal={IEEE Transactions on Neural Networks and Learning Systems}, 
  title={Two-Stage {B}ayesian Optimization for Scalable Inference in State-Space Models}, 
  year={2022},
  volume={33},
  number={10},
  pages={5138-5149},
  doi={10.1109/TNNLS.2021.3069172}}

@article{Doucet:Statistics:2003,
author = {Doucet, Arnaud
 and Tadić, Vladislav B.},
title = {Parameter estimation in general state-space models using particle methods
},
volume = {55},
journal = {Annals of the Institute of Statistical Mathematics},
pages = {409 -- 422},
year = {2003},

}

@article{Ionides:PNAS:2006,
author = {E. L. Ionides  and C. Bretó  and A. A. King },
title = {Inference for nonlinear dynamical systems},
journal = {Proceedings of the National Academy of Sciences},
volume = {103},
number = {49},
pages = {18438-18443},
year = {2006},
doi = {10.1073/pnas.0603181103},
URL = {https://www.pnas.org/doi/abs/10.1073/pnas.0603181103},
eprint = {https://www.pnas.org/doi/pdf/10.1073/pnas.0603181103},
}

@INPROCEEDINGS{Poyiadjis:ACC:2006,
  author={Poyiadjis, G. and Singh, S.S. and Doucet, A.},
  booktitle={American Control Conference}, 
  title={Gradient-free maximum likelihood parameter estimation with particle filters}, 
  year={2006},
  volume={},
  number={},
  pages={6},
  doi={10.1109/ACC.2006.1657187}}

@article{COURTS:Auto:2023,
title = {Variational system identification for nonlinear state-space models},
journal = {Automatica},
volume = {147},
pages = {110687},
year = {2023},
issn = {0005-1098},
doi = {https://doi.org/10.1016/j.automatica.2022.110687},
url = {https://www.sciencedirect.com/science/article/pii/S0005109822005519},
author = {Jarrad Courts and Adrian G. Wills and Thomas B. Schön and Brett Ninness}
}

@ARTICLE{Pitt:Warwick:2002,
  author={Michael K Pitt },
  journal={Warwick Economic Research Paper}, 
  title={Smooth Particle Filters for Likelihood Evaluation and Maximisation}, 
  year={2002},
  number={651}}

@article{Pitt:JASA:1999,
 ISSN = {01621459},
 URL = {http://www.jstor.org/stable/2670179},
 author = {Michael K. Pitt and Neil Shephard},
 journal = {Journal of the American Statistical Association},
 number = {446},
 pages = {590--599},
 publisher = {[American Statistical Association, Taylor & Francis, Ltd.]},
 title = {Filtering via Simulation: Auxiliary Particle Filters},
 urldate = {2024-05-11},
 volume = {94},
 year = {1999}
}

@book{ljung:book:1999,
  title={System Identification: Theory for the User},
  author={Ljung, L.},
  isbn={9780136566953},
  lccn={98018554},
  year={1999},
  publisher={Prentice Hall PTR}
}

@article{Johansen:SC:2008,
 author = {Johansen, Adam M. and Doucet, Arnaud and Davy, Manuel},
 journal = {Statistics and Computing},
 number = {},
 pages = {47-57},
 title = {Particle methods for maximum likelihood estimation in latent variable models},
 volume = {18},
 year = {2008}
}

@article{Andrieu:JRSS:2010,
    author = {Andrieu, Christophe and Doucet, Arnaud and Holenstein, Roman},
    title = {Particle {M}arkov Chain {M}onte {C}arlo Methods},
    journal = {Journal of the Royal Statistical Society Series B: Statistical Methodology},
    volume = {72},
    number = {3},
    pages = {269-342},
    year = {2010},
    month = {05},
    issn = {1369-7412},
    doi = {10.1111/j.1467-9868.2009.00736.x},
    url = {https://doi.org/10.1111/j.1467-9868.2009.00736.x},
    eprint = {https://academic.oup.com/jrsssb/article-pdf/72/3/269/49515565/jrsssb\_72\_3\_269.pdf},
}

@article{Mingas:IJAR:2017,
title = {Particle {MCMC} algorithms and architectures for accelerating inference in state-space models},
journal = {International Journal of Approximate Reasoning},
volume = {83},
pages = {413-433},
year = {2017},
issn = {0888-613X},
doi = {https://doi.org/10.1016/j.ijar.2016.10.011},
url = {https://www.sciencedirect.com/science/article/pii/S0888613X16302092},
author = {Grigorios Mingas and Leonardo Bottolo and Christos-Savvas Bouganis}
}

@book{cappé:book:2006,
  title={Inference in Hidden {M}arkov Models},
  author={Capp{\'e}, O. and Moulines, E. and Ryden, T.},
  isbn={9780387289823},
  lccn={2005923551},
  year={2006},
  publisher={Springer New York}
}

@book{doucet:Book:2001,
  title={{S}equential {M}onte {C}arlo Methods in Practice},
  author={Doucet, A. and Smith, A. and de Freitas, N. and Gordon, N.},
  isbn={9780387951461},
  lccn={00047093},
  year={2001},
  publisher={Springer New York}
}

@book{elliott:book:1995,
  title={{H}idden {M}arkov Models: Estimation and Control},
  author={Elliott, R.J. and Aggoun, L. and Moore, J.B.},
  isbn={9780387943640},
  lccn={94028643},
  year={1995},
  publisher={Springer New York}
}

@book{west:book:1999,
  title={Bayesian Forecasting and Dynamic Models},
  author={West, M. and Harrison, J.},
  isbn={9780387947259},
  lccn={96038166},
  year={1999},
  publisher={Springer New York}
}

@article{Picchini:CS:2018,
title = {Coupling stochastic {EM} and approximate {B}ayesian computation for parameter inference in state-space models},
journal = {Computational Statistics},
volume = {33},
pages = {179-212},
year = {2018},
author = {Picchini, Umberto and Samson, Adeline}
}

@article{Wills:Auto:2013,
title = {Identification of {H}ammerstein–{W}iener models},
journal = {Automatica},
volume = {49},
number = {1},
pages = {70-81},
year = {2013},
issn = {0005-1098},
author = {Adrian Wills and Thomas B. Schön and Lennart Ljung and Brett Ninness},
}

@article{PITT:JE:2012,
title = {On some properties of {M}arkov chain {M}onte {C}arlo simulation methods based on the particle filter},
journal = {Journal of Econometrics},
volume = {171},
number = {2},
pages = {134-151},
year = {2012},
issn = {0304-4076},
doi = {https://doi.org/10.1016/j.jeconom.2012.06.004},
url = {https://www.sciencedirect.com/science/article/pii/S0304407612001510},
author = {Michael K. Pitt and Ralph dos Santos Silva and Paolo Giordani and Robert Kohn}
}

@ARTICLE{Bobak:IEEE:2016,
  author={Shahriari, Bobak and Swersky, Kevin and Wang, Ziyu and Adams, Ryan P. and de Freitas, Nando},
  journal={Proceedings of the IEEE}, 
  title={Taking the Human Out of the Loop: A Review of {B}ayesian Optimization}, 
  year={2016},
  volume={104},
  number={1},
  pages={148-175},
  doi={10.1109/JPROC.2015.2494218}}

@article{Tian:JES:2020,
title = {One-shot parameter identification of the {T}hevenin’s model for batteries: Methods and validation},
journal = {Journal of Energy Storage},
volume = {29},
pages = {101282},
year = {2020},
issn = {2352-152X},
doi = {https://doi.org/10.1016/j.est.2020.101282},
author = {Ning Tian and Yebin Wang and Jian Chen and Huazhen Fang},
}

@ARTICLE{Tian:TCST:2020,
  author={Tian, Ning and Fang, Huazhen and Chen, Jian and Wang, Yebin},
  journal={IEEE Transactions on Control Systems Technology}, 
  title={Nonlinear Double-Capacitor Model for Rechargeable Batteries: Modeling, Identification, and Validation}, 
  year={2021},
  volume={29},
  number={1},
  pages={370-384},
  doi={10.1109/TCST.2020.2976036}}

@ARTICLE{Sitterly:TSE:2011,
  author={Sitterly, Mark and Wang, Le Yi and Yin, G. George and Wang, Caisheng},
  journal={IEEE Transactions on Sustainable Energy}, 
  title={Enhanced Identification of Battery Models for Real-Time Battery Management}, 
  year={2011},
  volume={2},
  number={3},
  pages={300-308},
  doi={10.1109/TSTE.2011.2116813}}

@ARTICLE{Yu:TIE:2017,
  author={Yu, Zhihao and Xiao, Linjing and Li, Hongyu and Zhu, Xuli and Huai, Ruituo},
  journal={IEEE Transactions on Industrial Electronics}, 
  title={Model Parameter Identification for Lithium Batteries Using the Coevolutionary Particle Swarm Optimization Method}, 
  year={2017},
  volume={64},
  number={7},
  pages={5690-5700},
}

@article{Goshtasbi:JPS:2024,
title = {Enhanced equivalent circuit model for high current discharge of lithium-ion batteries with application to electric vertical takeoff and landing aircraft},
journal = {Journal of Power Sources},
volume = {620},
pages = {235188},
year = {2024},
issn = {0378-7753},
doi = {https://doi.org/10.1016/j.jpowsour.2024.235188},
url = {https://www.sciencedirect.com/science/article/pii/S0378775324011406},
author = {Alireza Goshtasbi and Ruxiu Zhao and Ruiting Wang and Sangwoo Han and Wenting Ma and Jeremy Neubauer}
}

@book{Sarkka:Cambridge:2023,
  title = {Bayesian Filtering and Smoothing}, 
  publisher = {Cambridge University Press},
  author = {S\"{a}rkk\"{a},  Simo and Svensson,  Lennart},
  year = {2023},
  edition = {2}
}

@BOOK{Plett:2015,
  title     = {Battery Management Systems: Equivalent-Circuit Methods Volume {II}},
  author    = "Plett, Gregory",
  publisher = "Artech House", 
  edition   =  2, 
  year      =  2015,  
}

@article{Ahmed:SAE:2015,
  title = {Model-Based Parameter Identification of Healthy and Aged {Li-ion} Batteries for Electric Vehicle Applications},
  volume = {4}, 
  number = {2},
  journal = {SAE International Journal of Alternative Powertrains}, 
  author = {Ahmed,  Ryan and Gazzarri,  Javier and Onori,  Simona and Habibi,  Saeid and Jackey,  Robyn and Rzemien,  Kevin and Tjong,  Jimi and LeSage,  Jonathan},
  year = {2015}, 
  pages = {233–247}
}

@article{Lagarias:SIAM:1998,
author = {Lagarias, Jeffrey C. and Reeds, James A. and Wright, Margaret H. and Wright, Paul E.},
title = {Convergence Properties of the Nelder--Mead Simplex Method in Low Dimensions},
journal = {SIAM Journal on Optimization},
volume = {9},
number = {1},
pages = {112--147},
year = {1998}, 
}
\end{document}